\documentstyle[eqsecnum,preprint,prd,aps]{revtex}

\def\OO#1{{\cal O}(c^{-#1})}
\def\ve#1{{\bf #1}}

\begin{document}

\draft

\title{Relativistic Celestial Mechanics with PPN Parameters}

\author{Sergei A. Klioner and Michael H. Soffel}
\address{
           Lohrmann Observatory,
           Dresden Technical University,\\
           Mommsenstra\ss e 13,
           D-01062 Dresden, Germany\\
}

\date{10 June 1999}

\maketitle

\begin{abstract}
Starting from the global parametrized post-Newtonian (PPN) reference
system with two PPN parameters $\gamma$ and $\beta$ we consider
a space-bounded subsystem of matter and construct a local reference
system for that subsystem in which the influence of external masses
reduces to tidal effects. Both the metric tensor of the local PPN
reference system in the first post-Newtonian approximation as well as
the coordinate transformations between the global PPN reference system
and the local one are constructed in explicit form. The terms
proportional to $\eta=4\beta-\gamma-3$ reflecting a violation of
the equivalence principle are discussed in detail.

We suggest an empirical definition of multipole moments which are
intended to play the same role in PPN celestial mechanics as the
Blanchet-Damour moments in General Relativity. We also show that
the tidal gravitational field as seen in the local PPN reference
system can be expanded into powers of local coordinates similarly to
the tidal expansion in General Relativity.

Starting with the metric tensor in the local PPN reference system we
derive translational equations of motion of a test particle (an Earth
satellite) in that system. The translational and rotational equations
of motion for center of mass and spin of each of $N$ extended massive
bodies possessing arbitrary multipole structure are derived. All
equations of motion are presented also in the form of multipole
expansions. Several interesting features of the equations are
discussed.

As an application of the general equations of motion a monopole-spin
dipole model is considered and the known PPN equations of motion of
mass monopoles with spins are rederived. For the first time, these
equations are derived in a self-consistent manner which does not
require any additional assumptions about the behavior of bodies such as
secular stationarity.
\end{abstract}

\pacs{PACS numbers: 04.25.Nx, 04.80.Cc, 95.10.Ce}

%%% preprint
\newpage

%%% preprint
\tableofcontents

%%% preprint
\newpage

\section{Introduction}
\label{Introduction}

Einstein's theory of gravity (the so-called General Theory of
Relativity, GRT) in the last several decades has evolved far from being
predominantly a playground for mathematicians.  Not only that the field
of experimental gravity has expanded and the number of experimental
tests of GRT increased drastically, but certain results from GRT are
taken for granted and implemented in software for solving
technologically oriented problems. A good example is Very Long Baseline
Interferometry (VLBI) that determines baselines on the surface of the
Earth, and the position of the rotation pole with subcentimeter accuracy,
the length of the day at the fraction of a millisecond level, and
relative angles between two remote radio sources with a precision
better than a milliarcsecond.  In the VLBI analysis software
relativistic effects like the gravitational time delay are routinely
taken into account. Note that the gravitational light deflection
amounts to $1.75''$ at the limb of the Sun and it decreases with the
inverse impact parameter of the (unperturbed) light-ray from the Sun.
This implies that the light deflection (more precisely the
gravitational time delay) in the field of the Sun is measurable even
for radio sources lying closely to $180^\circ$ from the Sun
\cite{will:93,soffel:89}.

For the analysis of high precision optical astrometric measurements at
the milliarcsecond level (e.g., HIPPARCOS) GRT became an indispensable
tool \cite{Froeschle:Mignard:Arenou:1997}.
For planned space interferometry with microarcsecond
accuracies (DIVA, GAIA, SIM, FAME, etc.) relativity will play a
crucial role and the whole process of observation (essentially,
following the photons from the emission region onto the CCD-sensor)
has to be formulated within the framework of Einstein's theory of
gravity \cite{Klioner:Kopeikin:1992,Lindegren:Perryman:1996}.

Astrometry is one important field of application of GRT, relativistic
astrophysics (e.g., the physics of white dwarf stars, neutron stars,
black  holes or quasars, gravitational wave physics, cosmology) and
celestial mechanics are further ones.  Techniques used to gather
information about the dynamics of astronomical bodies comprise  the
timing of millisecond pulsars at the sub-microsecond level, laser
ranging to selected artificial satellites such as LAGEOS and to
retro corner reflectors that have been placed on the lunar surface
at the centimeter
level and radar ranging to planets or spacecrafts (with, e.g., a few
meters accuracy for the Viking landers on Mars).

Meanwhile it is well recognized that GRT enters every astronomical
observation or measuring technique at a certain level of accuracy. Such
a critical level has been reached for a variety of geodetic
measurements including GPS and a long time ago already for the
procedure to establish a terrestrial time scale (TT, TAI).
In 1991 the International Astronomical Union and the International
Union for Geodesy and Geophysics adopted recommendations implying the
use of GRT in modeling of modern high-accuracy observations
\cite{iau:91}. The International Earth Rotation Service
responsible for precise monitoring of the Earth orientation
parameters has also adopted relativistic models for several kinds of
astronomical observations \cite{iers:96}.

For practical applications one faces the problem of solving Einstein's
field equations for complicated situations. Methods of numerical
relativity might be employed in the future not only for problems
related with gravity wave generation (e.g., for the description of
coalescing binaries) or cosmology, but also for celestial mechanical
problems in the solar system. However, the {\it classical} formulation
of celestial mechanics is simplified considerably by the introduction
of collective variables like mass-multipole moments, center of mass,
etc. and one clearly wants to keep these advantages in a relativistic
formulation.  Note that the use of such collective variables is
related with expansions in terms of a 'geometric coupling parameter'
$\alpha \equiv L/R\,$ , measuring the ratio between a characteristic
linear dimension, $L$, of the bodies and a characteristic separation,
$R$, between them.

Already soon after Einstein published his first papers on his theory
of gravity, Einstein \cite{ein:15}, Droste \cite{dro:16}, De Sitter
\cite{sit:16}, and Lorentz and Droste \cite{lor:17} devised an
approximation method (called ``post-Newtonian'') which allowed them to
compare General Relativity with Newton's theory of gravity, and to
predict several ``relativistic effects'' in celestial mechanics, such
as the relativistic advance of the perihelion of planets, and the
relativistic precession of the lunar orbit: this post-Newtonian
approach to general relativistic celestial mechanics was subsequently
developed by many authors, notably by Fock \cite{foc:55}, Papapetrou
\cite{pap:51}, Chandrasekhar and colleagues
\cite{cha:65,cha:69,cha:70}, Caporali \cite{cap:81}, Grishchuk and
Kopejkin \cite{gri:86} and many others (for a review of the
development of the problem of motion in General Relativity see, e.g.,
Damour \cite{dam:87}).  Basically the post-Newtonian approach is a
slow-motion and weak-field approximation to Einstein's theory of
gravity. One introduces one or several dimensionless parameters for
the $N$-body problem: $\beta_e \equiv v_{\rm orb}/c$ (orbital
velocities), $\beta_i \equiv v_{\rm int}/c$ (internal or intrinsic
rotational velocities), $\kappa_e \equiv GM/(c^2 R)$ (external
gravitational potential) and $\kappa_i \equiv GM/(c^2 L)$ (internal
gravitational potential) if $M$ is the mass of a body with dimension
$L$ and $R$ is a typical distance to other bodies. Usually in the
post-Newtonian approximation one assumes $\beta_i^2 < \beta_e^2 \sim
\kappa_e < \kappa_i = \epsilon \ll 1 \,$.  E.g., $\epsilon < 10^{-5}$
everywhere in the solar system.

All of the applications mentioned above require a correspondingly
accurate relativistic theory of the gravitational $N$-body problem
consisting of

\medskip
(i) the external problem: to determine the motion of the centers of
mass of the $N$ bodies and

\smallskip
(ii) the internal problem: to determine the motion of each body around
its center of mass.

\medskip

It is a remarkable fact of Newtonian celestial mechanics that these two
sub-problems are coupled only very weakly. The first reason is the fact
that the ellipticities (dimensionless mass-quadrupole moments or
$J_2$-values) and higher mass-multipole moments of solar system bodies
are all very small. Therefore the motions of the center of masses are
dominated by the (approximately constant) masses of the bodies
independent of internal motion, and the couplings of higher
mass-multipole moments to higher-order derivatives of the external
gravitational potential (i.e., couplings to the external curvature
tensor in a geometric language) are very small. On the other hand the
local environment of a body is influenced by external bodies only
through the action of {\it tidal forces} that grow linearly with the
distance from the body's center of mass and vary with $L$ according to
$(R/L)^3$. This remarkable decoupling is sometimes expressed by the
phrase ``effacement of internal degrees of freedom in the global problem
and of the external world in the local system''.

A second reason for this effacement is the principle of equivalence. In
its weak form, applicable to Newtonian celestial mechanics, it asserts
the equivalence of inertial mass with gravitational mass that has been
tested by experiments with a relative precision of about $10^{-11}$
(see e.g., Will \cite{will:93} for more detail).

As explained in great detail, e.g., in
\cite{brum:kopej:89a,kopej:88,dsx:91}, the treatments of both
sub-problems in the classical post-Newtonian approach are
unsatisfactory. One usually considers the $N$-body problem in some
asymptotically flat space-time that can be covered with one single
global coordinate system, $x^\mu \equiv (ct,x,y,z) \equiv (ct,x^i)$, $i =
1,2,3$. In the classical formulation of post-Newtonian theory concepts
like ''center of mass'', ''mass-multipole moments'' (including the
mass) or ''mass-centered coordinates'' are defined in this global
coordinate system. E.g., for applications in the solar system the
global coordinates usually will be barycentric ones and geocentric
spatial coordinates classically were defined by $X^i = x^i - x_E^i(t)
\, ,$ where $x_E^i(t)$ denotes the global coordinates of the
''geocenter''. Such definitions do not define a useful mass-centered
frame in General Relativity, in the sense that it does not efface the
external gravitational field down to tidal effects. In such a
geocentric coordinate system relativistic effects are of order
$\beta_e^2 \sim 10^{-8}$ and not of order $\kappa_i = (GM_E/c^2 R_E)
\approx 7 \cdot 10^{-10}$ as one would expect.  Clearly, the classical
post-Newtonian theory tries to fit GRT into purely Newtonian framework
and bears traces of Newton's absolute space and time.

In the recent decade the relativistic theory of astronomical reference
systems in the framework of General Relativity has been considered in a
detailed manner by a number of authors
\cite{ashby:bertotti:84,ashby:bertotti:86,brum:kopej:89a,kopej:88,%
brum:kopej:89b,kopej:91,dsx:91,dsx:92,dsx:93,klio:voi:93}.  These
investigations were motivated both by theoretical interest to construct
a proper reference system for a massive material subsystem and by practical
requirements of astronomy \cite{koval:muell:81,seidel:85,guinot:86}
dictated by increased accuracy levels.
The two most advanced and complete approaches to construct
a physically adequate local reference system in General Relativity are
the Brumberg-Kopeikin approach
\cite{brum:kopej:89a,kopej:88,brum:kopej:89b,kopej:91,klio:voi:93} and
the Damour-Soffel-Xu one \cite{dsx:91,dsx:92,dsx:93,dsx:94}.
For the gravitational $N$-body problem both formalisms introduce a
total of $N+1$ different coordinate systems: one set of global
coordinates $x^\mu = (ct,x^i)$ and one set of local coordinates
$X^\alpha = (cT, X^a)$ for each body comoving with the body under
consideration. The relation between global and local (e.g., geocentric)
coordinates describing one and the same event is assumed to be of
the general form

\begin{equation}
x^\mu(X^\alpha) = x_{\rm E}^\mu(T) + e^\mu_a(T) X^a + \xi^\mu(T,\ve{X}),
\end{equation}

\noindent
where $\xi^\mu$ is at least quadratic in the local space coordinate
$X^a$ and $x^\mu_E(T)$ describes the motion of a central worldline of the
Earth (later chosen as the center of mass).  Equivalently this
transformation is described by

\begin{equation}
X^\alpha(x^\mu) = X_{\rm E}^\alpha(t) + E^\alpha_i(t) r_E^i +
\Xi^\alpha(t,\ve{r}_E)\, ,
\end{equation}

\noindent
with $r^i_E(t) = x^i - x_E^i(t)$ and $\Xi^\alpha$ is
assumed to be at least quadratic in $r_E^i$. Note that in passing from
one form to the other one has to bear in mind that $x_E^i(T)$ and
$X_E^a(t)$ describe two different points on the central geocentric
worldline \cite{brum:kopej:89b,dsx:94}. Suitable effacement conditions
in the $X^\alpha$ system lead to constraints on the functions
appearing in the coordinate transformations. Note, that these
transformations involve space and time coordinates; they present
generalized Lorentz transformations. Only by such transformations
apparent Lorentz-contraction effects etc. can be avoided in the
geocentric coordinate system.

Principle features and results of the theory of local reference system
in the framework of General Relativity are:

\smallskip
-- a description of the metric tensor $g_{\mu\nu}$ in the global system
with only two gravitational potentials $(w,w^i)$ as functions of $x^\mu$;
similarly, a description of $G_{\alpha\beta}$ in some local system with
only two potentials $(W,W^a)$ \cite{dsx:91};

\smallskip
-- a new theory of astronomical reference frames, including the
transformation rules for coordinates
\cite{brum:kopej:89a,kopej:88,dsx:91,dsx:92,klio:voi:93}
and metric potentials \cite{dsx:92};

\smallskip
- an improved description, with full post-Newtonian accuracy, of the
gravitational structure of each body by means of a set of multipole
moments, $(M_L^A, S_L^A)$, of a body $A$ (e.g., the Earth) which are
linked in an operational way to what can be observed in the local
gravitational environment of $A$ \cite{dsx:91,dsx:92};

\smallskip
- a description of the influence of the external world in a local frame
by means of some suitably defined tidal moments
\cite{brum:kopej:89a,kopej:88,dsx:91,dsx:92,klio:voi:93};

\smallskip
- translational \cite{dsx:91,dsx:92} and
rotational \cite{dsx:93} equations of motion with full
post-Newtonian accuracy and inclusion of all multipole moments;

\smallskip
- physical adequate equations of motion of a test particle (e.g.,
Earth satellite) in the local reference system
\cite{brum:kopej:89b,dsx:94,klio:voi:93};

\smallskip
- physically adequate relativistic models for many kinds of
observations (VLBI
\cite{Kopeikin:1990,Soffel:Wu:Xu:Mueller:1991,Klioner:1991},
high-accuracy positional observations
\cite{brum:kopej:89a,Klioner:Kopeikin:1992}, various methods of remote
clock synchronization \cite{Klioner:1992}, etc.)

\bigskip
Despite the fact that GRT so far passed all experimental tests with
flying colours (e.g., Misner {\it et al.}\cite{mtw}, Will \cite{will:93})
people never lost interest in alternative theories of gravity. This is
one reason why a {\it parametrized post-Newtonian formalism} (the PPN
formalism) was developed to cover the post Newtonian limits of a whole
class of metric theories of gravity including GRT.  A primitive version
of such a formalism was devised by Eddington \cite{edd:22} and
Robertson \cite{rob:62} who introduced the space curvature parameter
$\gamma$ and the non-linearity parameter $\beta$ that are both equal to
one ($\beta = \gamma = 1$) in GRT, but may take different values in
alternative theories of gravity. Later Schiff \cite{sch:67} included
terms from the rotation of bodies (gravito-magnetic terms). The
classical PPN-formalism was developed mainly by Kenneth Nordtvedt in
the late 60s and early 70s in collaboration with Clifford Will
(Nordtvedt\cite{nor:68,nor:70}, Nordtvedt and Will\cite{nor:72}, Will
and Nordtvedt\cite{wil:72}). More PPN-parameters have been added to
cover the post-Newtonian limits of a variety of scalar-tensor theories
like the Brans-Dicke theory, vector tensor theories, bimetric theories,
etc. The choice of PPN-parameters has changed over the years until some
standard was achieved that can be found in Will\cite{will:93}.  Here, a
total of ten PPN-parameters is introduced not only to distinguish
between the various post-Newtonian limits of different theories of
gravity, but also to provide a tool to characterize the precision of
relevant gravitational experiments. In the following we will restrict
ourselves to the Eddington-Robertson parameters $\beta$ and $\gamma$.
The other PPN-parameters that are related with possible preferred
locations, preferred frames or a violation of conservation of total
momentum will not be considered here. This is not a crucial
restriction for our approach that can be extended in a straightforward
manner to any choice of PPN parameters. However, considering that
the two parameters are the most important ones we confine ourselves to
$\beta$ and $\gamma$ only.

This Eddington-Robertson-Schiff Nordtvedt-Will PPN-formalism is
formulated in the spirit of classical post-Newtonian framework
described above.  Nevertheless, it was highly successful and lead to
new predictions, e.g., related with a breakdown of the strong
equivalence principle.  Let us summarize briefly the most interesting
results from the classical PPN-formalism that one would expect to be
valid even in an improved version.

\medskip
1. In every metric theory of gravity a test body of negligible
extension and negligible gravitational self-energy will follow a
geodesic in a suitably defined external metric $g_{\mu\nu}^{\rm ext}$.
From the usual form of the PPN metric one finds that the geodetic
acceleration of a body $E$ is given by

\begin{eqnarray}
a_{E,{\rm geodesic}}^i = -\sum_{B \not= E} G\,M_B\,{r_{EB}^i \over r_{EB}^3}
&+& {1 \over c^2} \left( \dots \right)
\end{eqnarray}

\noindent
with suitably chosen masses $M_B$ of gravitating bodies and
$r_{EB}^i=x_E^i-x_B^i$. Note that
this acceleration (or the geodetic equation) follows from an action
integral for the worldline of some central point of body $E$ of the
form

\begin{equation}
I^E_{\rm geodesic} = - M_E\,c\,\int
\left[ - g^{\rm ext}_{\mu\nu} dx^\mu dx^\nu \right]^{1/2}.
\end{equation}

\noindent
The corresponding $N$-body Lagrangian takes the form \cite{will:93}
($\ve{r}_{AB}=\ve{x}_A-\ve{x}_B$)

\begin{eqnarray}
{\cal L}^{\rm N\,\,body}_{\rm geodesic} = &-&\sum_{A} M_A \, c^2\,\left(
1 - \frac 12 {\ve{v}_A^2 \over c^2} - \frac 18 {\ve{v}_A^4 \over c^4} \right)
\nonumber\\
&+&{1\over 2}\sum_A\sum_{B\not=A}
{G M_A M_B \over r_{AB}}\,
\Biggl[ 1 + (2\gamma + 1){\ve{v}_A^2 \over c^2} - (2\beta-1)
\sum_{C\not=A} {G M_C \over c^2\, r_{AC} }
\nonumber \\
&-&{1\over 2c^2} (4\gamma+3)\, \ve{v}_A \cdot \ve{v}_B
- {1\over 2c^2}\,
(\ve{v}_A \cdot \ve{r}_{AB})\,
(\ve{v}_B \cdot \ve{r}_{AB})\,{1\over r_{AB}^2}\,
\Biggr].
\end{eqnarray}

\medskip
2.  In general a body $E$ with non negligible gravitational
self-energy $\Omega_E$ will not follow a geodesic because of a
violation of the Strong Equivalence Principle (e.g.,
Nordtvedt\cite{nor:68,nor:70}, Will\cite{will:93}). In that case the
value for the gravitational mass of $E$, $M_E^{(G)}$, will differ
from that for the inertial mass $M_E$ by

\begin{equation}
M_E^{(G)} = M_E + \eta\,\Omega_E/c^2 \,.
\end{equation}

\noindent
Here, $\eta$ is the Nordtvedt-parameter. A classical result gives its
relation to the Eddington-Robertson parameters

\begin{equation}
\eta = 4 \beta - \gamma - 3\,.
\end{equation}

\noindent
This is the well-known Nordtvedt effect leading to a polarization of
the lunar orbit around the Earth in the direction of the Sun if $\eta
\not= 0$.  The ratio of gravitational to inertial mass has an
interesting and valuable interpretation. A violation of the
Strong Equivalence
Principle implies that in the local (e.g. geocentric) system the
effects from the outside world cannot be effaced down to the usual
tidal forces. Instead the ``effective''
gravitational constant, $G_{\rm local}$, as experienced
in the local $E$-frame becomes effectively space dependent

\begin{equation}
G_{\rm local} = G \left( 1 - \eta {\overline{U} \over c^2}\right) \,.
\end{equation}

\noindent
Here, $\overline{U}$ is the external gravitational potential produced by the
outside world, i.e.~by all gravitating bodies different from $E$.
Since second and higher order derivatives of $\overline{U}$ give rise to the
usual tidal forces the effacement breaks down for the zeroth and first
derivative of $\overline{U}$, taken at the geocenter $\ve{x}_E$.
Considering the gravitational action between two points $\ve{x}$ and
$\ve{x}'$ one has effectively \cite{Nordtvedt:1998}

\begin{equation}\label{G-local-linear}
G_{\rm local} = G
\left[ 1 - \eta {\overline{U}(\ve{x}_E) \over c^2} - \frac \eta 2 \, \ve{a}_E
\cdot( \ve{x} + \ve{x}') \right] \, ,
\end{equation}

\noindent
where we have replaced $\nabla \overline{U}(\ve{x}_E)$ by the
acceleration $\ve{a}_E$.  Deriving the acceleration, e.g., of a
satellite in the vicinity of the Earth by classical means, but with
this expression for $G_{\rm local}$ leads to the above ratio of
gravitational to inertial mass of the Earth.

\medskip
3. Already in Newtonian celestial mechanics extended bodies in general
do not experience just the $1/r^2$ acceleration when interacting with
another body at distance $r$. Instead higher-multipole mass moments
($C_{lm}, S_{lm}$ or equivalently $M_L$ in the Cartesian language  with
$l \ge 2$) lead to a deviation of the 'geodetic' free-fall behavior.
To lowest order this force of  geodetic-deviation is given by

\begin{equation}\label{ai-quarupole-Newt}
a_{E,{\rm quadrupole}}^i =
{1\over 2\,M_E} \, M^E_{jk} \, \partial_{ijk}\, \overline{U}(\ve{x}_E) =
{1\over 2\,M_E} \, I^E_{jk} \, \partial_{ijk}\, \overline{U}(\ve{x}_E).
\end{equation}

\noindent
Here,

\begin{equation}\label{Ijk-Newt}
I^E_{jk} = \int_E \rho^* r_E^j r_E^k d^3x
\end{equation}

\noindent
is the Newtonian second-order mass tensor related with the Newtonian
trace-free quadrupole tensor, $M^E_{jk}$, by

\begin{equation}\label{Mjk=I<jk>-Newt}
M^E_{jk} = I^E_{jk} - \frac 13 I^E_{ss} \, \delta_{jk}.
\end{equation}

\noindent
Note that the last equality in (\ref{ai-quarupole-Newt})
follows from the Laplace equation in the absence of gravitational
sources. Since second spatial derivatives of the external potential can
be interpreted as components of a Newtonian external curvature tensor
this quadrupole acceleration of body $E$ can be interpreted as
resulting from a coupling of the body's second-moment mass tensor with
the curvature tensor produced by the outside world. For that reason in
any metric theory of gravity one expects the world line of a suitably
chosen central point inside body $E$ to be influenced by the body's
covariant second-moment mass tensor, $I^E_{\alpha\beta}$, via a
coupling to the external curvature tensor $R^{\rm ext}_{\alpha\mu\beta\nu}$ as
described by the action integral (Nordtvedt \cite{Nordtvedt:1994})

\begin{equation}
I = - M_E\,c\,\int \left[ - \left(g^{\rm ext}_{\mu\nu} + {I^{\alpha\beta}_E \over M_E}
R^{\rm ext}_{\alpha\mu\beta\nu} \right) dx^\mu dx^\nu \right]^{1/2}.
\end{equation}

\noindent
In the Newtonian limit only the spatial components of
$I^{\alpha\beta}_E$ (as measured in the local geocentric frame)
contribute
and from the action integral one derives an acceleration due to the
second-moment mass tensor of

\begin{equation}
a_{E,{\rm quadrupole}}^i =
\frac {c^2}{2\,M_E}\, I^E_{jk}\,\partial_i R^{\rm ext}_{j0k0}
= \frac 1{2\,M_E}\, I^E_{jk}\,
\partial_{ijk} \overline{U}(\ve{x}_E) \,.
\end{equation}

\noindent
As in Newton's theory, in GRT one finds that the trace of $I^E_{jk}$
does not lead to an additional acceleration because of the vacuum
field equation (i.e., the vanishing of the Ricci-tensor in source-free
regions).  However, in general for $\gamma \not=1$ the Ricci-tensor in
vacuum does not vanish in alternative theories of gravity. In that
case there will be an additional acceleration due to the
extension-curvature coupling proportional to $\gamma - 1$ even for
spherical bodies with

\begin{equation}
I^E_{jk} = \frac 13 I^E \delta_{jk}
\end{equation}

\noindent
and

\begin{equation}
I^E \equiv \int_E \rho^* r_E^i r_E^i d^3x \,.
\end{equation}

\noindent
This implies that the effacement in GRT in the local (geocentric)
system does not only lead to the ordinary value of the gravitational
constant (since $\eta = 0$), but also prevents extension-curvature
couplings for spherical bodies (since $\gamma = 1$).

\medskip
4. Finally problems related with the spin (intrinsic angular momentum)
have been treated extensively in the literature. There are two aspects
that have to be tackled: (i) the influence of the spin upon the
translational equation of motion and (ii) the motion of the spin vector
itself (the rotational motion).  Progress for solving the first problem
has been achieved by Mathisson \cite{mat:37}, Papapetrou\cite{pap:51},
Corinaldesi and Papapetrou\cite{cor:51} without definitive conclusive
results for GRT. However, in the first post-Newtonian approximation the
situation is different. The spin-orbit and spin-spin coupling terms
have been derived by Brumberg \cite{brum:72} who also discusses the
physical implications in detail (see also Kalitzin \cite{kal:59} and
Michalska \cite{mic:60}).  They have also been derived using a
one-graviton exchange theory by Barker {\it et al.} \cite{bar:75} and
by B\"orner {\it et al.} \cite{boe:1975} with parameters $\beta$ and
$\gamma$ (see also Barker {\it et al.},
\cite{bar:76,bar:81,bar:82,bar:86}). Tulczyjew \cite{Tulczyjew:1959}
has derived formally the same results using distributional sources. As first
discussed by Damour \cite{Damour:1982}, if we require invariance under
Lorentz boosts, the spin-orbit Lanrangian depends not only on positions
and velocities but also on accelerations. Within the Lorentz-invariant
Lagrangian formalism this is the manifestation of the well-known Thomas
precession \cite{Damour:Taylor:1992}.  The Lagrangian for the
spin-orbit and spin-spin coupling terms for body $E$ takes the form
\cite{Damour:Taylor:1992,bar:75,soffel:89}

\begin{eqnarray}\label{Lagr:spin}
{\cal L}^E_{\rm spin} =
&+&{1\over c^2}\,(1+\gamma)\,G\,\sum_{B\not=E}
{\Bigl( (\ve{v}_E-\ve{v}_B) \times \ve{r}_{EB} \Bigr) \cdot
\left(M_E\,\ve{S}_B+M_B\,\ve{S}_E\right)
\over r^3_{EB}}
\nonumber\\
&+&{1\over 2c^2}\,({\ve{v}}_E \times \ve{a}_E) \cdot \ve{S}_E
\nonumber\\
&-&{1\over 2c^2}\,(1+\gamma)\,G\, \sum_{B \not= E}
\left[ {3\, (\ve{S}_E \cdot \ve{r}_{EB})\,(\ve{S}_B \cdot \ve{r}_{EB} ) \over
r_{EB}^5 } - {\ve{S}_E \cdot \ve{S}_B \over r_{EB}^3 } \right] \,.
\end{eqnarray}

\noindent
Already to Newtonian order the motion of the spin is governed by
extension-curvature couplings (precession of the equinoxes; e.g.,
Misner {\it et al.} \cite{mtw}), were the quadrupole and higher order
mass-multipole moments couple to the external curvature tensor. In the
absence of such tidal forces the spin vector is Fermi-Walker
transported (e.g., Misner {\it et al.} \cite{mtw}) along the body's
central world line. This Fermi-Walker transport has been extensively
discussed in the literature. One finds that the spin-vector in the
comoving frame precesses with respect to remote celestial objects (more
precisely with respect to the local coordinate induced tetrad, see e.g.
\cite{will:93,soffel:89}), i.e.

\begin{equation}
{d\ve{S} \over d\tau} = \ve{\Omega} \times \ve{S}
\end{equation}

\noindent
with

\begin{equation}
\ve{\Omega} = - \frac 12 {\ve{v} \times \ve{Q}  \over c^2} - \frac 12
{\nabla \times \ve{W} \over c^2}
+ \left( \gamma + \frac 12 \right) {\ve{v} \times \nabla U \over c^2} \,.
\end{equation}

\noindent
Here, $\ve{v}$ is the global coordinate velocity of the rotating body
(gyroscope), $\ve{Q}$ are the spatial components of the body's
four-acceleration and $\ve{W}$ the three space-time components of the
metric tensor in the comoving system. The three terms on the right hand
side describe the well known Thomas precession (first term), Schiff or
Lense-Thirring precession (second term) and the geodetic or de
Sitter-Fokker precession (third term).  The extension-curvature coupling
in the post-Newtonian framework of GRT have been discussed in detail in
\cite{dsx:93}. For recent articles on this coupling in GRT see e.g.,
Apostolatos \cite{apo:96a,apo:96b}.

\bigskip

It is the purpose of this article to extend and improve the classical
PPN framework with parameters $\beta$ and $\gamma$ so to cover also the
principal results from the Brumberg-Kopeikin and DSX formalisms.

Both Brumberg-Kopeikin and DSX approaches use the field equations of
GRT in order to construct the local reference system. However, in the
PPN formalism we do not have any kind of ``parametrized post-Newtonian
field equations'', but only the metric tensor in the global reference
system. To construct a physically adequate local reference system means
to find a suitable coordinate transformation which maps the global
barycentric metric tensor into a ``good'' local one having some
desired properties. These desired properties in General Relativity read

\begin{itemize}

\item[{\bf A.}] The gravitational field of external bodies is
represented only in the form of a relativistic tidal potential which is at
least of second order in the local spatial coordinates and coincides
with the usual Newtonian tidal potential in the Newtonian limit;

\item[{\bf B.}] The internal gravitational field of the subsystem
coincides with the gravitational field of a corresponding isolated
source provided that the tidal influence of the external matter is
neglected.

\end{itemize}

These two requirements can simultaneously be satisfied in General
Relativity as has been shown in the framework of the Brumberg-Kopeikin
and DSX formalisms. It is clear that this fact is closely related with
the validity of the Strong Equivalence Principle in GRT. In the PPN
formalism due to a possible violation of the Strong Equivalence
Principle one expects that either property {\bf A} or {\bf B} can be
satisfied, but not both of them simultaneously.

Both Brumberg-Kopeikin and DSX formalisms rely essentially upon the
field equations of general relativity. In contrast to this here we do
not want to confine ourselves to a particular theory of gravity and use
instead the phenomenological framework of PPN formalism.  It is clear
that considering a particular theory of gravity might provide a deeper
insight into the physical origin of various effects in that particular
theory. However, this physical meaning would be valid only in that
theory of gravity and cannot be ``generalized'' onto other theories of
gravity even if they give formally the same post-Newtonian metric.

The global barycentric PPN metric is assumed to be a solution of the
field equations of any metric gravity theory covered by the PPN
formalism.  The construction of a physically adequate ("good") local
reference system involves a suitable coordinate transformation which
converts the global barycentric coordinates into "good" local ones.  It
is clear that if the original barycentric metric represents a solution
of the field equations, then the resulting geocentric metric is also a
solution of the same equations. The metric tensor of the local
reference system is sufficient to discuss both translational and
rotational equations of motion in that local reference system since
they follow either from the geodetic equations (for test bodies and
photons) or from the local equations of motion (for extended bodies).
Both kinds of equations require only the metric tensor for their
detailed evaluation. No other possible scalar, vector, etc. fields
appearing in a particular non-Einsteinian theory of gravity play a role
in discussing equations of motion in the PPN formalism.  This allows us
to construct the local reference system from some {\sl geometrical}
considerations. An attempt to achieve this has been already undertaken
by Shahid-Saless and Ashby \cite{sah:ashby:88}, who introduced a kind
of Fermi normal coordinates in a simplified model case of two
spherically symmetric nonrotating gravitating bodies.  However it is
clear that Fermi coordinates in the PPN framework lead to the same
principal difficulties when considering the local subsystem with a
complicated multipole structure, to which they lead in GRT (see, e.g.,
\cite{kopej:91,dsx:91}). We use the much more powerful and elegant
approach of the Brumberg-Kopeikin and DSX formalisms.

A practical application of the theory of relativistic
astronomical reference systems lies in the construction
physically adequate models for modern
high-precision astronomical observations (depending on the kind of
observations the models involve the
coordinate transformations between the reference systems, equations of
light propagation in both reference systems as well as equations of
translational and rotational motion of extended bodies).
However, on the other hand, these observations yield a
significant amount of experimental data that can be used for testing
metric theories of gravity such as General Relativity. The Parametrized
Post-Newtonian (PPN) formalism \cite{will:93} presently can be
considered to represent the standard framework for such tests in the
first post-Newtonian approximation.
Therefore, it is quite important to have a consistent
theory of relativistic astronomical reference systems in the PPN
formalism. Besides a purely theoretical interest to construct a
"good" proper reference system for a massive subsystem (a massive
body) in the framework of the PPN formalism, a PPN theory of
astronomical reference systems is also important from a rather
practical point of view, since presently there is a
contradiction between the recommendations of the International
Astronomical Union \cite{iau:91} concerning the relativistic reference
system to be used for data reduction which are valid only in GRT
and the common practice to estimate the PPN parameters (at least
$\gamma$ and $\beta$) from observations.  Moreover, the
International Earth Rotation Service being an official service of the
IAU recommends in its IERS Conventions \cite{iau:91} to estimate the
PPN parameters $\gamma$ and $\beta$ in routine data processing.

This paper focuses mainly on problems of reference systems and
celestial mechanics. More specifically, we present

\medskip
-- a new PPN theory of astronomical reference systems (Sections
\ref{Metric-tensors-and-coordinate-transformations}--
\ref{Nordtvedt-section});

\smallskip
-- a new description of the local gravitational environment of a body
with new PPN mass- and spin-multipole moments generalizing the
Blanchet-Damour moments from \cite{dsx:91} (Section
\ref{Section-multipoles-internal}) and of the tidal forces (Section
\ref{Section-multipoles-external});

\smallskip
-- new PPN satellite equations of motion (Section
\ref{section:eqm-test-particle}).

\smallskip
-- new translational and rotational global equations of motion of $N$
extended, rotating and gravitationally interacting bodies of arbitrary
shape and composition with all multipole moments of gravitating bodies
with parameters $\beta$ and $\gamma$ (Section
\ref{translational-eqm});

\smallskip

\noindent
A overall outlook and a summary of our main results are given in
Section \ref{Conclusions-and-outlook}.  As expected all results that
have been listed under 1-4 above are recovered, but now with the help
of a new and improved PPN-framework.  Some partial results have already
been published elsewhere
\cite{klio:97,klioner:soffel:1998,klioner:soffel:1999}.

\section{Notations}

Greek indices $\alpha, \beta, \dots, \mu, \nu, \dots$ running from 0 to
3 indicate all four space-time components of the corresponding
variable.  Latin indices $a,b,\dots, i,j,\dots$ run from 1 to 3 and
refer to three spatial components of the corresponding variable.  We
use Einstein's summation convention for both types of indices
independent of the position of repeated indices: e.g., $x^i\,x^i\equiv
(x^1)^2+(x^2)^2+(x^3)^2$.  We use several special objects:
$\delta^{ij}={\rm diag}(1,1,1)$ is the Kronecker delta;
$\varepsilon_{ijk}$ is the fully antisymmetic Levi-Civita symbol
($\varepsilon_{123}=+1$); $\eta_{\mu\nu}={\rm diag}(-1,+1,+1,+1)$ is
the Minkowki flat space metric (we use the signature $(-+++)$
throughout this paper).  Capital latin indices $A,B,C,\dots = 1 \dots
N$ (as subscripts or superscripts) refer to the bodies of our $N$-body
system: e.g., $\Phi_E$ is the value of some variable $\Phi$ related to
body $E$.

The global coordinates are designated by $x^\mu=(ct,x^i)$, $t$ being
coordinate time and $x^i$ being spatial coordinates of the global
reference system. The global metric is $g_{\mu\nu}(x^\lambda)$ and we
systematically use lower case latin letters for the quantities
belonging to the global reference system. As in the DSX papers
\cite{dsx:91,dsx:92,dsx:93,dsx:94} the space-time indices referring to
global coordinates are taken from the second part of the greek
alphabet $\mu,\nu,\dots$ and space indices are taken from the second
part of the latin alphabet $i,j,\dots$. The local coordinates for body
$E$ are designated by $X^\alpha=(cT,X^a)$, $T$ being coordinate time
and $X^a$ being spatial coordinates of the global reference
system. The local metric is $G_{\alpha\beta}(X^\gamma)$ (do not
confuse an index $\beta$ with the PPN parameter $\beta$). The
space-time indices referring to global coordinates are taken from the
first part of the greek alphabet $\alpha,\beta,\dots$ and space
indices are taken from the first part of the latin alphabet
$a,b,\dots$.

A comma before an index designates the partial derivative with respect
to the corresponding coordinates: $A_{,\mu}={\partial A(t,\ve{x})/\partial
x^\mu}$, $A_{,i}={\partial A(t,\ve{x})/\partial x^i}$,
$A_{,\alpha}={\partial A(T,\ve{X})/\partial X^\alpha}$, $A_{,a}={\partial
A(T,\ve{X})/\partial X^a}$. For partial derivatives with respect to the
coordinate times $t$ and $T$ we use $A_{,t}={\partial A(t,\ve{x})/\partial
t}$, $A_{,T}={\partial A(T,\ve{X})/\partial T}$. A semicolon before a
space-time index denotes the corresponding covariant derivative.
Parentheses surrounding a group of indices denote symmetrization, e.g.,
$A_{(ij)}={1\over 2}\left(A_{ij}+A_{ji}\right)$. Brackets surrounding
two indices denote antisymmetrization, e.g., $A_{i[jk]}={1\over
2}\left(A_{ijk}-A_{ikj}\right)$. Angle brackets surrounding a group of
indices or, alternatively, a caret on top of a tensor symbol denote the
symmetric trace-free (STF) part of the corresponding object, e.g.,
$\hat A_{ab} \equiv A_{\langle ab \rangle}\equiv STF_{ab} A_{ab}=
A_{(ab)}-{1\over 3}\,\delta^{ab}\,A_{cc}$.  We shall freely lower and
raise spatial indices, e.g. $A_a \equiv A^a$.  For sequences of spatial
indices we shall use multiindices in the same way as it is done, e.g.,
in \cite{dsx:91}: a spatial multi-index containing $l$ indices is
denoted by $L$ ($K$ for $k$ indices, etc.): $L=i_1\dots i_l$ if $L$
refers to global coordinates and $L=a_1\dots a_l$ in the case of local
ones. We use also $L-1=i_1\dots i_{l-1}$, etc. A multisummation is
understood for repeated multiindices: $A_L\,B_L \equiv \sum_{i_1\dots
i_l} A_{i_1\dots i_l}\,B_{i_1\dots i_l}$. For a spatial vector $v^i$ we
denote $v^L\equiv v^{i_1}\,v^{i_2}\dots v^{i_l}$.  For an $L$-order
partial derivative we denote $\partial_L\equiv
\partial_{i_1}\dots\partial_{i_l}$.

\section{Metric tensors and coordinate transformations}
\label{Metric-tensors-and-coordinate-transformations}

\subsection{The PPN metric tensor in the global reference system}

We consider a gravitational $N$ body system in an asymptotically
flat space-time that can be covered by one single (global) coordinate
system $x^\mu = (ct, x^i)$ with

\begin{equation}\label{BRS-limits}
\lim_{|\ve{x}|\rightarrow \infty \atop{t = {\rm const.}}} g_{\mu\nu} =
\eta_{\mu\nu},
\end{equation}

\noindent
where $g_{\mu\nu}$ is the metric tensor in the global coordinate system
and $\eta_{\mu\nu}={\rm diag}(-1,+1,+1,+1)$ is the flat metric tensor
of Minkowski space-time. The metric tensor $g_{\mu\nu}$ in the global
PPN reference system is written as

\begin{eqnarray}\label{BRS:metric}
g_{00}&=&-1+{2\over c^2}\,w(t,\ve{x})-{2\over c^4}
\,\beta\, w^2(t,\ve{x})+\OO5,
\nonumber \\
g_{0i}&=&-{2(1+\gamma)\over c^3}\,w^i(t,\ve{x})+\OO5,
\nonumber \\
g_{ij}&=&\delta_{ij}\left(1+{2\over c^2}\,\gamma\,
w(t,\ve{x})\right)+\OO4.
\end{eqnarray}

\noindent
In accordance with the classical PPN framework we will assume the
metric potentials $w$ and $w^i$ to obey the equations

\begin{eqnarray}\label{UW,ii-U,tt}
w_{,ii}-{1\over c^2}\,w_{,tt}=-4\,\pi\,G\,\sigma+\OO4,
\\ \label{Uijj}
w^i_{,jj}=-4\,\pi\,G\,\sigma^i+\OO2,
\end{eqnarray}

\noindent
where

\begin{eqnarray}\label{sigma}
\sigma&=&{1\over c^2}\,\left(T^{00}+\gamma\,T^{kk}+
{1\over c^2}\,T^{00}\,(3\gamma-2\beta-1)\,w\right)+\OO4,
\\ \label{sigmai}
\sigma^i&=&{1\over c}\,T^{0i}+\OO2.
\end{eqnarray}

\noindent
Here, $T^{\mu\nu}$ are the components of energy-momentum tensor in the
global reference system and $w$ in (\ref{sigma})is needed  only to
Newtonian order where it coincides with the Newtonian potential.
Because of requirement (\ref{BRS-limits}) $(w^\mu \equiv (w,w^i))$ we
have

\begin{equation}\label{w-mu-limits}
\lim_{|\ve{x}|\rightarrow \infty \atop{t = {\rm const.}}}
w^\mu(t,\ve{x}) = 0
\end{equation}

\noindent
and the solution of (\ref{UW,ii-U,tt})--(\ref{Uijj}) can be written in
the form

\begin{equation}\label{w-mu-solution}
w^\mu(t,\ve{x})= G \int {\sigma^\mu (t,\ve{x}') \over |\ve{x}-\ve{x}'|}d^3x'
+ {1\over 2 c^2}\,G\,{\partial^2 \over \partial t^2}\,
\int \sigma^\mu(t,\ve{x}') | \ve{x} - \ve{x}' | d^3x'+\OO4.
\end{equation}

The expressions above are valid independently of the particular form of
the energy-momentum tensor. This is no more than a formal way to
specify the global PPN metric tensor.  If matter is described as an ideal
fluid or
stressed continuum the metric (\ref{BRS:metric})--(\ref{w-mu-solution})
coincides with the version of the PPN formalism described in
\cite{will:93} as well as with the version discussed in \cite{mtw},
provided that only two parameters $\gamma$ and $\beta$ are retained in
both versions. Effectively we consider only those theories of
gravity which produce the metric (\ref{BRS:metric})--(\ref{sigmai})
in the first post-Newtonian approximation for some form of the
energy-momentum tensor.

Eqs. (\ref{UW,ii-U,tt})--(\ref{sigmai}) can be considered as
``effective'' or ``phenomenological'' post-Newtonian field equations in
the global reference system. These equations reduce to the Einstein
field equations for $\gamma=\beta=1$.  It is, however, clear that these
phenomenological field equations have nothing to do with the actual
field equations of a specific theory of gravity covered by the PPN
formalism.

The system under study is supposed to be spatially bounded and isolated
which means that $T^{\mu\nu}=0$ sufficiently far from the origin.  We suppose
also that the matter in concentrated in spatially separated blobs
(bodies) and $T^{\mu\nu}$ vanishes outside these blobs. One of
such blobs is selected to define a local subsystem and the matter
inside the blob is referred below to as internal matter. All other matter
is called external.  Any subsystem is characterized by a world tube $V$
which encompasses the internal matter. Since the subsystems are
supposed to be spatially separated we choose the world tube $V$ in such
a way that $T^{\mu\nu}=0$ on the boundary of $V$. In the following we will
concentrate upon one of these bodies that we will call ``Earth''
(indicated by subscript $E$).

The potentials $w^\mu$ are defined by (\ref{w-mu-solution}) as volume
integrals over the whole 3-space.  Therefore, splitting the area of
integration into the volume $V$ of the body, for which we want to
construct a local reference system, and the remaining part of space, we
split $w^\mu$ into internal potentials (potentials of the body under
consideration) and external ones (potentials due to the other bodies):

\begin{eqnarray}\label{BRS:split}
w(t,\ve{x})&=&w_{\rm E}(t,\ve{x})+{\overline{w}}(t,\ve{x}),
\nonumber \\
w^i(t,\ve{x})&=&w^i_{\rm E}(t,\ve{x})+{\overline{w}}^i(t,\ve{x}).
\end{eqnarray}

\noindent
The internal potentials $w_{\rm E}^\mu=(w_{\rm E},w_{\rm E}^i)$ are
defined as

\begin{equation}\label{wE-mu}
w_{\rm E}^\mu(t,\ve{x})= G \int_V {\sigma^\mu (t,\ve{x}') \over
|\ve{x}-\ve{x}'|}d^3x' + {1\over 2 c^2}\,G\,{\partial^2 \over \partial
t^2}\, \int_V \sigma^\mu(t,\ve{x}') | \ve{x} - \ve{x}' | d^3x'.
\end{equation}

We adopt a harmonic-like gauge for the global PPN metric tensor.
Precisely speaking, the global metric tensor satisfies the usual
harmonic gauge ($g={\rm det}(g_{\mu\nu})$)

\begin{equation}\label{harmonic-gauge}
{\partial\over \partial x^\alpha}
\left( {(-g)}^{1/2} g^{\alpha\beta} \right)=0
\end{equation}

\noindent
in case of General Relativity $\beta=\gamma=1$. This requires

\begin{equation}\label{w,t+wi,i=0}
w_{,t}+w^i_{,i}=\OO2,
\end{equation}

\noindent
which is also compatible with the definitions of $w$ and $w^i$, and
the Newtonian continuity equation

\begin{equation}\label{Newton-continuity}
\sigma_{,t}+\sigma^i_{,i}=\OO2.
\end{equation}

\noindent
Since the energy-momentum tensor $T^{\mu\nu}$ is supposed to vanish on
the boundary of the world tube $V$ encompassing the subsystem, Eq.
(\ref{Newton-continuity}) allows one to formulate (\ref{w,t+wi,i=0})
separately for internal and external potentials

\begin{equation}\label{wE,t+wEi,i=0}
w_{E,t}+w^i_{E,i}=\OO2,
\end{equation}

\begin{equation}\label{wext,t+wexti,i=0}
{\overline w}_{,t}+{\overline w}^i_{,i}=\OO2.
\end{equation}

The metric (\ref{BRS:metric})--(\ref{w-mu-solution}) is equivalent to
the PPN metric in \cite{will:93} up to a trivial gauge transformation.
The transformations between our reference system $(t,x^i)$ and the PPN
reference system in the standard post-Newtonian gauge $(t_{\rm PN},x^i_{\rm
PN})$ reads

\begin{eqnarray}\label{tx-tx-PPN}
t_{\rm PN}&=&t-{1\over c^4}\,\chi_{,t}+\OO5,
\nonumber \\
x^i_{\rm PN}&=&x^i,
\end{eqnarray}

\noindent
where $\chi$ is the superpotential, here defined by

\begin{equation}\label{chi}
\chi={1\over 2}\,G\,\int \sigma(t,\ve{x}')\,|\ve{x}-\ve{x}'|\,d^3x'+\OO2,
\end{equation}

\noindent
so that

\begin{equation}\label{chi,ii=w}
\chi_{,ii}=w+\OO2.
\end{equation}

The transformation of the results derived below into the standard PPN
gauge represents no principal difficulty and can be done in exactly
the same way as it has been done in Section V of \cite{klio:voi:93}.

\subsection{Energy-momentum tensor for an ideal fluid}

Although, as noted above, our formalism is not restricted to any
particular form of the energy-momentum tensor, the standard PPN formalism
(see, e.g., \cite{will:93}) is elaborated for the case of an ideal fluid and
it is interesting to specify our formalism for this
simplified case. The energy-momentum tensor of an ideal fluid
in the global asymptotically flat reference system $(t,x^i)$
reads

\begin{eqnarray}\label{Tmunu}
T^{00}&=&\rho c^2\left[1+{1\over c^2}
\left(v^2+\Pi+2U\right)\right]+\OO2,
\nonumber \\
T^{0i}&=&\rho c\, v^i\left[1+{1\over c^2}
\left(v^2+\Pi+2U\right)\right]+
{1\over c} p\,v^i+\OO3,
\nonumber \\
T^{ij}&=&\rho\, v^i\, v^j+\delta^{ij} p+\OO2.
\end{eqnarray}

\noindent
Here $v^i=v^i(t,\ve{x})$ is the velocity of matter,
$\Pi(t,\ve{x})$ is the specific energy density (ratio of energy
density to rest-mass density), $p(t,\ve{x})$ is the isotropic pressure,
$\rho(t,\ve{x})$ is the invariant density (that is, the density
invariant under coordinate transformations from $(t,\ve{x})$ to any
$(\tilde t,\tilde{\ve{x}})$:  $\rho(t,\ve{x})=\rho(\tilde
t,\tilde{\ve{x}})$). In many cases it is more convenient to
express the energy-momentum tensor in terms of
$\rho^*(t,\ve{x})$,

\begin{equation}\label{rho*}
\rho^*(t,\ve{x})=\rho(t,\ve{x})
(-g)^{1/2} {cdt\over ds}=\rho\left[1+{1\over c^2}
\left({1\over 2} v^2+3\gamma U\right)+\OO4\right].
\end{equation}

\noindent
For the density $\rho^*$ a Newtonian-like continuity equation
holds to post-Newtonian accuracy

\begin{equation}\label{rho*-cont}
{\partial \over \partial t} \rho^*+{\partial \over \partial x^i}
\left( \rho^* \, v^i \right)=\OO4.
\end{equation}

\noindent
In terms of $\rho^*$ the energy-momentum tensor reads

\begin{eqnarray}\label{Tmunu*}
T^{00}&=&\rho^* c^2\left[1+{1\over c^2}
\left({1\over 2}\,v^2+\Pi-(3\gamma-2)\,U\right)\right]+\OO2,
\nonumber \\
T^{0i}&=&\rho^* c\,v^i\,\left[1+{1\over c^2}\left(
{1\over 2}\,v^2+\Pi-
(3\gamma-2)\,U\right)\right]+
{1\over c}\,p\,v^i+\OO3,
\nonumber \\
T^{ij}&=&\rho^*\,v^i\,v^j+\delta^{ij}\,p+\OO2.
\end{eqnarray}

\noindent
Here and above $U$ is the Newtonian-like gravitational potential

\begin{equation}\label{U}
U(t,\ve{x})=G\,\int {\rho^*(t,\ve{x}')\over |\ve{x}-\ve{x}'|} d^3x'.
\end{equation}

For an ideal fluid the potentials $w$ and $w^i$ can then be expressed as

\begin{eqnarray}
\label{w}
w&=&U+{1\over c^2}\left(
\left(\gamma+{1\over 2}\right)
\Phi_1 + (1-2\beta)\Phi_2 + \Phi_3 + 3\gamma \Phi_4 + \chi_{,tt}\right),
\\ \label{wi}
w^i&=&G\,\int {\rho^*(t,\ve{x}')\,v^i(t,\ve{x}')
\over|\ve{x}-\ve{x}'|}\,d^3x'+\OO2,
\\ \label{Phi1}
\Phi_1&=&G\,\int {\rho^*(t,\ve{x}') \dot x'^2 \over |\ve{x}-\ve{x}'|} d^3x',
\\ \label{Phi2}
\Phi_2&=&G\,\int {\rho^*(t,\ve{x}') U(t,\ve{x}') \over |\ve{x}-\ve{x}'|} d^3x',
\\ \label{Phi3}
\Phi_3&=&G\,\int {\rho^*(t,\ve{x}') \Pi(t,\ve{x}')
\over |\ve{x}-\ve{x}'|} d^3x',
\\ \label{Phi4}
\Phi_4&=&G\,\int {p(t,\ve{x}') \over |\ve{x}-\ve{x}'|} d^3x'.
\end{eqnarray}

Since all the potentials (\ref{U}), (\ref{Phi1})--(\ref{Phi4}) and
(\ref{chi}) are again some volume integrals, we can split them into
local parts where the integration is performed over a world tube $V$
covering the subsystem for which the local RS is being constructed
and external ones (the integration is performed only outside of
$V$)

\begin{eqnarray}\label{BRS:split:individual}
U&=&U_{\rm E}+{\overline{U}},
\nonumber \\
\Phi_i&=&\Phi_i^{\rm E}+{\overline\Phi}_i,\ i=1,2,3,4,
\nonumber \\
\chi&=&\chi^{\rm E}+\overline{\chi}.
\end{eqnarray}

\noindent
For example, the definition of $U_E$ reads

\begin{equation}\label{UE}
U_{\rm E}(t,\ve{x})=G\,\int_V {\rho^*(t,\ve{x}')\over |\ve{x}-\ve{x}'|} d^3x'.
\end{equation}

\subsection{The PPN metric tensor in the local geocentric system}

  We assume that the metric tensor $G_{\mu\nu}(T,\ve{X})$ in the local
reference system $X^\mu=(cT,X^a)$ has the same functional form as the
metric in the global reference system

\begin{eqnarray}\label{GRS:metric}
G_{00}&=&-1+{2\over c^2}\,W(T,\ve{X})-{2\over c^4}
\,\beta\,W^2(T,\ve{X})+{\cal O}(c^{-5}),
\nonumber \\
G_{0a}&=&-{2(1+\gamma)\over c^3}\,W^a(T,\ve{X})+{\cal O}(c^{-5}),
\nonumber \\
G_{ab}&=&\delta_{ab}\left(1+{2\over c^2}\,\gamma\,
W(T,\ve{X})\right)+{\cal O}(c^{-4}),
\end{eqnarray}

\noindent
and the potentials $W$ and $W^i$ entering into (\ref{GRS:metric}) admit
a representation in the form

\begin{eqnarray}\label{GRS:split}\label{W-split}
W(T,\ve{X})&=&W_{\rm E}(T,\ve{X})+Q_a(T)\,X^a+W_{\rm T}(T,\ve{X})
+{1\over c^2}\,\Psi(T,\ve{X}),
\\ \label{Wi-split}
W^a(T,\ve{X})&=&W^a_{\rm E}(T,\ve{X})
+{1\over 2}\,\varepsilon_{abc}\,C_b(T)\,X^c
+W^a_{\rm T}(T,\ve{X}).
\end{eqnarray}

\noindent
Here $W_T$ and $W_T^a$ are external potentials representing tidal
fields of the other bodies of the system and are assumed to be ${\cal
O}(\ve{X}^2)$. The two arbitrary functions $Q_a$ and $C_a$ have a clear
physical meaning which will be discussed below.  The local internal
gravitational potentials are represented by three other functions
$W_{\rm E}$, $W_{\rm E}^a$  and $\Psi$.
The potentials $W_{\rm E}$ and $W_{\rm E}^a$ are supposed
to have the same functional form as their counterparts $w_{\rm E}$ and
$w_{\rm E}^i$, but expressed in local coordinates:

\begin{eqnarray}
\label{WE}
W_{\rm E}&=&G\,\int_V\,\Sigma(T,\ve{X}')\,{1\over |\ve{X}-\ve{X}'|}\,d^3X'
\nonumber\\
&&
+{1\over 2c^2}\,G\,{\partial^2\over\partial T^2}\,
\int_V\,\Sigma(T,\ve{X}')\,|\ve{X}-\ve{X}'|\,d^3X'+{\cal O}(c^{-4}),
\\ \label{WEi}
W^a_{\rm E}&=&G\,\int_V\,\Sigma^a(T,\ve{X}')\,
{1\over |\ve{X}-\ve{X}'|}\,d^3X'
+{\cal O}(c^{-2}),
\end{eqnarray}

\begin{eqnarray}\label{Sigma}
\Sigma&=&
{1\over c^2}\left({\cal T}^{00}+\gamma\, {\cal T}^{aa}+
{1\over c^2}\, {\cal T}^{00}\,(3\gamma-2\beta-1)\,W\right)+{\cal O}(c^{-4}),
\\ \label{Sigmai}
\Sigma^a&=&{1\over c}\, {\cal T}^{0a}+{\cal O}(c^{-2}),
\end{eqnarray}

\noindent
where ${\cal T}^{\mu\nu}$ is the energy-momentum tensor in the local
reference system $(T,X^a)$. Finally, the function $\Psi(T,\ve{X})$
represents any possible deviation of the actual internal gravitational
field as seen in the local reference system from the form
(\ref{WE})--(\ref{WEi}). Clearly the appearance of $\Psi$ is related
with a violation of the Strong Equivalence Principle which makes it
impossible to satisfy simultaneously properties {\bf A} and {\bf B}
formulated in Section \ref{Introduction}. By assuming that $W_{\rm T}$
and $W^a_{\rm T}$ are $\sim {\cal O}(\ve{X}^2)$ we assume that
property {\bf A} is satisfied. For that reason property {\bf B} will
be violated resulting in the appearance of the additional quantity
$\Psi$.

The local metric is also supposed to satisfy the harmonic gauge (${\cal
G}={\rm det}(G_{\alpha\beta})$)

\begin{equation}\label{harmonic-gauge-GRS}
{\partial\over \partial X^\alpha}
\left( {(-{\cal G})}^{1/2} G^{\alpha\beta} \right)=0
\end{equation}

\noindent
in case of General Relativity ($\beta=\gamma=1$). This requires

\begin{equation}\label{W,T+Wa,a=0}
W_{,T}+W^a_{\ ,a}=\OO2.
\end{equation}

\noindent
From (\ref{WE}) and (\ref{WEi}), the Newtonian continuity equation

\begin{equation}\label{Sigma-continuity}
\Sigma_{,T} + \Sigma^a_{\ ,a} = \OO2
\end{equation}

\noindent
and from the assumption that the energy-momentum tensor (and,
therefore, $\Sigma$ and $\Sigma^a$) vanishes on the boundary of the
world tube $V$ it follows

\begin{equation}\label{WE-hamonicity}
W_{{\rm E},T}+W^a_{{\rm E},a}=\OO2.
\end{equation}

\noindent
From (\ref{W,T+Wa,a=0}) and (\ref{WE-hamonicity}) one finds

\begin{equation}\label{WT,u+Qi,uXi+WTi,i=0}
\dot Q_a X^a+W_{{\rm T},T}+W^a_{{\rm T},a}=\OO2.
\end{equation}

Let us note that at this point there is no way to prove that a local
reference system with all these properties exists. Unlike the case of
General Relativity we do not have field equations that would allow us
to prove some properties of the local reference system. We just
formulate the desired properties of the local reference system and then
prove that it is indeed possible to construct such a system.

\subsection{Coordinate transformations between the global and
local reference systems}

The results of the Brumberg-Kopeikin and DSX formalisms (see,
e.g., Theorems 1 and 2 of \cite{dsx:91}) allow us to
write the transformations between the global and local reference
systems in the form

\begin{eqnarray}
\label{trans:time}
&&T=t-{1\over c^2}\left(A+v_{\rm E}^ir_{\rm E}^i\right)+
{1\over c^4}\left(B+B^i r_{\rm E}^i+ B^{ij} r_{\rm E}^i r_{\rm E}^j
+C(t,\ve{x})
\right)+{\cal O}(c^{-5}),
\\ \label{trans:space}
&&X^a=R^a_{\,j}\left(r_{\rm E}^j+{1\over c^2}
\left({1\over 2}v_{\rm E}^j\,v_{\rm E}^k r_{\rm E}^k+D^{jk}\,r_{\rm E}^k
+D^{jkl}\,r_{\rm E}^k r_{\rm E}^l\right)\right)
+{\cal O}(c^{-4}),
\end{eqnarray}

\noindent
where $r_{\rm E}^i=x^i-x_{\rm E}^i(t)$, $x_{\rm E}^i(t)$ is the
coordinates of the origin of the local reference system relative to the
global one, and $v_{\rm E}^i=dx_{\rm E}^i/dt$ and $a_{\rm
E}^i=d^2x_{\rm E}^i/dt^2$ are its velocity and acceleration,
respectively. The functions $A(t)$, $B(t)$, $B^i(t)$, $B^{ij}(t)$,
$D^{ij}(t)$, $D^{ijk}(t)$, $R^a_{\ j}(t)$ (being a rotational
(orthogonal) matrix) and
$C(t,\ve{x})\sim{\cal O}(r_{\rm E}^3)$ are some unknown functions to
be determined with the aid of matching of the two metric tensors.

The fact that the metric (\ref{BRS:metric}) can be transformed into
the metric (\ref{GRS:metric}) with the help of the transformations
(\ref{trans:time})--(\ref{trans:space}) is not obvious from the
beginning and has to be verified.

\section{Matching of the global and local PPN metric tensors}
\label{Matching-of-the-global-and-local-PPN-metric-tensors}

Matching of the metric tensors (\ref{BRS:metric}) and (\ref{GRS:metric})

\begin{equation}\label{matching}
g_{\varepsilon\lambda}(t,\ve{x})=
{\partial X^\mu\over \partial x^\varepsilon}\,
{\partial X^\nu\over \partial x^\lambda}\,
G_{\mu\nu}(T,\ve{X}),
\end{equation}

\noindent
where the Jacobian $\partial X^\mu / \partial x^\varepsilon$ has to be
determined from the coordinate transformation
(\ref{trans:time})--(\ref{trans:space}), yields the explicit
expressions for the unknown functions in the coordinate transformations
as well as in the local metric. The matching is performed in  four
steps, considering

\begin{itemize}
\item[A)] terms of order $\OO2$ in $g_{00}$
($\varepsilon=\lambda=0$ in (\ref{matching}));
\item[B)] terms of order $\OO2$ in $g_{ij}$
($\varepsilon=i$, $\lambda=j$ in (\ref{matching}));
\item[C)] terms of order $\OO3$ in $g_{0i}$
($\varepsilon=i$, $\lambda=0$ in (\ref{matching}));
\item[D)] terms of order $\OO4$ in $g_{00}$
($\varepsilon=\lambda=0$ in (\ref{matching})).
\end{itemize}

\noindent
At each step of the matching procedure
we equate separately four kind of terms in
the left- and right-hand sides of (\ref{matching}):

\noindent
\begin{itemize}
\item[1)] terms containing the internal potentials of either global or local
metric;
\item[2)] external terms independent of local spatial coordinates (that is,
the terms which are functions of time only);
\item[3)] external terms linear with respect to local coordinates $\ve{X}$;
\item[4)] external terms of at least second order with respect to local
coordinates $\ve{X}$.
\end{itemize}

\noindent
This procedure allows us to derive the following results.

\subsection{Matching of order  $\OO2$ in $g_{00}$}

\noindent
1) Internal terms give the Newtonian transformation between the
internal potentials in the global and local reference systems

\begin{equation}\label{WE-wE-Newt}
W_{\rm E}(T,\ve{X})=w_{\rm E}(t,\ve{x})+\OO2.
\end{equation}

\noindent
It is easy to see that (\ref{WE-wE-Newt}) is in agreement with the
definitions of $w_E(t,\ve{x})$ and $W_E(u,\ve{w})$ (see, (\ref{wE-mu})
and (\ref{WE})).

\bigskip\noindent
2) External terms of order $|\ve{X}|^0$ give

\begin{equation}\label{A}
{d\over dt}\,A = {1\over 2}v_E^2+{\overline{w}}(\ve{x}_E).
\end{equation}

\noindent
Here and below for any function $\Phi(t,\ve{x})$ we use
the shorthand notation $\Phi(\ve{x}_E) \equiv \Phi(t,\ve{x}_E(t))$.

\bigskip\noindent
3) External terms of order $|\ve{X}|^1$ lead to the
Newtonian equations of motion of the center of mass
of the local system relative to the global reference system

\begin{equation}\label{aE-Newt}
a_E^i(t)={\overline{w}}_{,i}(\ve{x}_E)-R^a_{\ i}\,Q_a+\OO2.
\end{equation}

\bigskip\noindent
4) External terms of order ${|\ve{X}|}^l$, $l\ge 2$
yield the expression for the external tidal potential
in the local metric

\begin{eqnarray}\label{W_T-Newt}
W_{\rm T}(T,\ve{X})&=&
{\overline{w}}(t,\ve{x})
-{\overline{w}}(\ve{x}_E)
-{\overline{w}}_{,i}(\ve{x}_E)\,r_E^i+\OO2.
\end{eqnarray}

\noindent
Here and below for any function $\Phi(t,\ve{x})$ the shorthand notation
$\Phi_{,i}(\ve{x}_E)$ denotes $ \partial \Phi(t,\ve{x}) / \partial x^i
\,$ evaluated at $\ve{x}=\ve{x}_E(t)$.

\subsection{Matching of order  $\OO2$ in $g_{ij}$}

\noindent
1) Internal terms are matched automatically due to (\ref{WE-wE-Newt}).

\bigskip\noindent
2) External terms of order $|\ve{X}|^0$ give

\begin{equation}\label{D^{ij}}
D^{ij}(t)=\delta^{ij}\,\gamma\,{\overline{w}}(\ve{x}_E).
\end{equation}

\bigskip\noindent
3) External terms of order $|\ve{X}|^1$ give

\begin{equation}\label{D^{ijk}}
D^{ijk}(t)={1\over 2}\,\gamma\,\left(\delta^{ij}a_E^k+
\delta^{ik}a_E^j-\delta^{jk}a_E^i\right).
\end{equation}

\noindent
Let us note here that this expression for $D^{ijk}$ follows from the
isotropic form of $g_{ij}$ and $G_{ab}$ and its relation to $g_{00}$
and $G_{00}$, respectively (this was proven by Theorem 2 of
\cite{dsx:91} in the case $\gamma=1$).

\bigskip\noindent
4) External terms of order ${|\ve{X}|}^l$, $l\ge 2$ are matched
automatically due to (\ref{W_T-Newt}).

\subsection{Matching of order $\OO3$ in $g_{0i}$}
\label{OO3-in-g0i}

\noindent
1) Equating internal terms results in the expression for the
local vector potential

\begin{equation}\label{W_E^a}
W_E^a(T,\ve{X})=R^a_{\ i}\,
\left(w_E^i(t,\ve{x})-v_E^i\,w_E(t,\ve{x})\right)+\OO2.
\end{equation}

\noindent
This agrees with the definitions (\ref{wE-mu}) and (\ref{WEi}) of
$w_E^i(t,\ve{x})$ and $W_E^a(T,\ve{X})$ and the relation

\begin{eqnarray}\label{Sigma-a-sigma-i}
\Sigma^a=R^a_{\ i}\,\left(\sigma^i-v_E^i\,\sigma\right) +\OO2,
\end{eqnarray}

\noindent
which can be derived the definitions (\ref{Sigmai}),
(\ref{sigma})--(\ref{sigmai}) of $\Sigma^a$, $\sigma^i$ and $\sigma$
with the help of the coordinate transformations
(\ref{trans:time})--(\ref{trans:space}).

\bigskip\noindent
2) External terms of order $|\ve{X}|^0$ require

\begin{equation}\label{B^i}
B^i(t)=-{1\over 2}\,v_E^2\,v^i_E+
        2(1+\gamma)\,{\overline{w}}^i(\ve{x}_E)-
        (2\gamma+1)\,v_E^i\,{\overline{w}}(\ve{x}_E).
\end{equation}

\bigskip\noindent
3) External terms of order $|\ve{X}|^1$ lead to

\begin{eqnarray}\label{B^ij}
B^{ij}(t)&=& -v_E^{(i}\,R^a_{\,j)}\,Q^{a}+
(1+\gamma){\overline{w}}^{(i,j)}(\ve{x}_E)
%\nonumber \\
%&&
-\gamma\,v_E^{(i}\,{\overline{w}}^{,j)}+{1\over 2}\,\gamma
\,\delta^{ij}\,{\dot{\overline{w}}}(\ve{x}_E),
\end{eqnarray}

\begin{eqnarray} \label{Rij-Ca}
c^2\,R^a_{\ i}\,\dot R^a_{\ j}&=&-(1+\gamma)\,\varepsilon_{ijk} R^a_{\ k}C_a
\nonumber \\
&&
-2(1+\gamma)\,\overline{w}^{[i,j]}(\ve{x}_{\rm E})
+(1+2\gamma)\,v_{\rm E}^{[i}\,\overline{w}_{,j]}(\ve{x}_{\rm E})
+v_{\rm E}^{[i}R^a_{\ j]}Q_a+{\cal O}(c^{-2}).
\end{eqnarray}

\noindent
Eq. (\ref{Rij-Ca}) relates the two functions $C_a$ and $R^a_{\ i}$.  These
functions together describe the spatial orientation of the local reference
system with respect to the global one.  According to (\ref{Rij-Ca}) one
can choose $C_a(T)$ so that $R^a_{\ i}=\delta^a_{\ i}$ and the
resulting local reference system does not rotate relative to the global
one, i.e., the local reference system is kinematically non-rotating.
Another possible choice $C_a=0$ results in a dynamically
nonrotating local reference system, and the orthogonal matrix $R^a_{\
i}$ in the transformations of the spatial coordinates represents the
well-known  de Sitter, Lense-Thirring and Thomas precessions.  Note
that (\ref{Rij-Ca}) contains the well-known results for Lense-Thirring and de
Sitter precessions in the PPN formalism (the second and third terms on
the right-hand side, respectively). On the other hand, the fourth term
represents Thomas precession appearing in special relativity and
therefore, must be independent of the PPN parameters, which is indeed
the case. The first terms in $B^i$ and $B^{ij}$ (Eqs.  (\ref{B^i}) and
(\ref{B^ij})) come also from special relativity theory and are also
independent of $\beta$ and $\gamma$.

\bigskip\noindent
4) External terms of order ${|\ve{X}|}^l$, $l\ge 2$ give
the external tidal vector potential in the local reference
system

\begin{eqnarray}\label{W_T^i}
W_{\rm T}^a(T,\ve{X})=&&R^a_{\ i}\,\biggl\{{\overline{w}}^i(t,\ve{x})-
{\overline{w}}^i(\ve{x}_{\rm E})
-{\overline{w}}^i_{,j}(\ve{x}_{\rm E})r_{\rm E}^j
-v_{\rm E}^i W_{\rm T}(T,\ve{X})
\nonumber \\
&&\phantom{R^a_{\ i}\,\biggl\{}
+{1\over 2(1+\gamma)}
\left(\dot D^{ijk}\,r_E^j\,r_E^k-C_{,i}\right)
\biggr\}+{\cal O}(c^{-2}).
\end{eqnarray}

\noindent
Eq. (\ref{W_T^i}) together with (\ref{wext,t+wexti,i=0}) and
(\ref{WT,u+Qi,uXi+WTi,i=0}) leads to a Poisson-type equation for
the unknown function $C(t,\ve{x})$

\begin{equation}\label{C:laplace}
C_{,ii}= 2\dot D^{iik}r_E^k-2(1+\gamma)\,\dot a_E^k r_E^k=
(\gamma-2)\,{\dot{a}_E}^k r_E^k.
\end{equation}

\noindent
This equation does not fix $C(t,\ve{x})$ uniquely, but only
up to a harmonic function. This fact reflects some ambiguity in
constructing the local reference system in the form
(\ref{GRS:metric})--(\ref{Wi-split}). Assuming the function
$C$ together with its first and second partial derivatives with respect to
$X^a$ to be equal to zero at $X^a=0$, the general solution of
(\ref{C:laplace}) reads

\begin{equation}\label{C:general}
C=C_1+C_2,
\end{equation}

\noindent
where

\begin{equation}\label{C_1}
C_1(t,\ve{x})={\gamma-2\over 10}\,r_E^2\,({\dot{a}_E}^i\,r_E^i)
\end{equation}

\noindent
is the particular solution of (\ref{C:laplace}), and

\begin{equation}\label{C_2}
C_2(t,\ve{x})=\sum_{p=3}^\infty{1\over p!}V_{a_1\dots\, a_p}
r_E^{a_1}\dots\, r_E^{a_p}.
\end{equation}

\noindent
Here $V_{a_1\dots\, a_p}(t), p\geq 3$ is a family of arbitrary STF
tensors. Any closed-form solution of (\ref{C:laplace}) leads to
a closed-form metric tensor of the local reference system. E.g., we
might choose the simplest solution
$C=C_1=(\gamma-2)\,r_E^2\,({\dot{a}_E}^i\,r_E^i)/10$
corresponding to $V_{a_1\dots\,
a_p}(t)\equiv 0$. In this case the tidal vector-potential reads

\begin{eqnarray}\label{W_T^i:simplest}
W_{\rm T}^a(T,\ve{X})=&&R^a_{\ i}\,\biggl\{{\overline{w}}^i(t,\ve{x})-
{\overline{w}}^i(\ve{x}_{\rm E})
-{\overline{w}}^i_{,j}(\ve{x}_{\rm E})r_{\rm E}^j
-v_{\rm E}^i\, W_{\rm T}(T,\ve{X})
\nonumber \\
&&\phantom{R^a_{\ i}\,\biggl\{}
+{2\gamma+1\over 5\,(1+\gamma)}\,r_E^i\,(\dot a_E^k\,r_E^k)
-{3\gamma-1\over 10\,(1+\gamma)}{\dot{a}_E}^i\,\ve{r}_E^2
\biggl\}
+{\cal O}(c^{-2}).
\end{eqnarray}

\noindent
Note once again that the factor $\gamma+1$ in denominators of
(\ref{W_T^i}) and (\ref{W_T^i:simplest}) cancels with the same one in
the numerator of (\ref{GRS:metric}) and no rational functions of
$\gamma$ appear in the metric tensor.

It is easy to see from (\ref{W_T^i})--(\ref{C:laplace}) that
independent of $C_2$ the tidal vector potential $W_T^i$ satisfies the
equation

\begin{equation}\label{WTa,bb}
W_{T,bb}^a={1-\gamma\over 1+\gamma}\,R^a_{\ i}\,\dot a^i_E+\OO2.
\end{equation}

\subsection{Matching of order $\OO4$ in $g_{00}$}

\noindent
1)  Matching of both terms $\OO2$ and $\OO4$ in $g_{00}$ gives

\begin{eqnarray}\label{g00:c2+c4:matching}
W_{\rm E}(T,\ve{X})+{1\over c^2}\,\Psi(T,\ve{X})
&=&
w_{\rm E}(t,\ve{x})\,\left(1+{1\over c^2}\,(\gamma+1)\,v_E^2\right)
\nonumber \\
&&-{2(\gamma+1)\over c^2} w_{\rm E}^i(t,\ve{x}) v_E^i
\nonumber \\
&&-{2(\beta-1)\over c^2} w_{\rm E}(t,\ve{x})
\,\left(\overline{w}(\ve{x}_E)+a_E^i r_E^i\right)+\OO4,
\end{eqnarray}

\noindent
generalizing (\ref{WE-wE-Newt}) to post-Newtonian accuracy.
On the other hand, one can relate the local internal potential $W_{\rm E}$
with the global ones $w_{\rm E}$ and $w_{\rm E}^i$ directly by
using their definitions
(\ref{wE-mu}) and (\ref{WE}). To this end, from the definitions
(\ref{Sigma}), (\ref{sigma})--(\ref{sigmai}) of $\Sigma$, $\sigma^i$
and $\sigma$, and the coordinate transformations
(\ref{trans:time})--(\ref{trans:space}) one derives

\begin{eqnarray}\label{Sigma-sigma}
\Sigma(T,\ve{X})&=&\sigma(t,\ve{x})\,
\left(1+{1\over c^2}\left((1+\gamma) v_E^2
+(2\beta-3\gamma-1)\,(\overline{w}(\ve{x}_E)+a_E^k\,r_E^k)\right)\right)
\nonumber\\
&&-{1\over c^2}\,2(1+\gamma)\,v_E^i\,\sigma^i+\OO4.
\end{eqnarray}

\noindent
Using this the following result generalizing Theorem 5 of \cite{dsx:91}
is found

\begin{eqnarray}\label{We-we}
W_{\rm E}(T,\ve{X})&=&
w_{\rm E}(t,\ve{x})\,\left(1+{1\over c^2}\,(\gamma+1)\,v_E^2\right)
\nonumber \\
&&-{2(\gamma+1)\over c^2} w_{\rm E}^i(t,\ve{x}) v_E^i
\nonumber \\
&&+{1\over c^2}\,(2\beta-\gamma-1)\, w_{\rm E}(t,\ve{x})
\,\left(\overline{w}(\ve{x}_E)+a_E^i r_E^i\right)
\nonumber \\
&&-{1\over c^2}\,(4\beta-\gamma-3)\,\chi^{\rm E}_{,i}\,a_E^i
+\OO4.
\end{eqnarray}

\noindent
Here,

\begin{equation}\label{chiE-bary}
\chi^{\rm E}(t,\ve{x})={1\over 2}\,G\,\int_V \sigma(t,\ve{x}')\,|\ve{x}-\ve{x}'|\,d^3x'+\OO2
\end{equation}

\noindent
and

\begin{equation}\label{chiE-geo}
\chi^{\rm E}(T,\ve{X})={1\over 2}\,G\,\int_V \Sigma(T,\ve{X}')\,|\ve{X}-\ve{X}'|\,d^3X'+\OO2=
\chi^{\rm E}(t,\ve{x})+\OO2.
\end{equation}

\noindent
A comparison of (\ref{g00:c2+c4:matching}) with (\ref{We-we}) shows that

\begin{eqnarray}\label{Psi}
\Psi(T,\ve{X})&=&-\eta \left(w_{\rm E}\,\left({\overline{w}}(\ve{x}_E)+
a_E^i\,r_E^i\right)
-\chi^{\rm E}_{,i}\,a_E^i\right)+\OO2,
\end{eqnarray}

\noindent
where

\begin{equation}\label{eta}
\eta=4\beta-\gamma-3
\end{equation}

\noindent
is the well-known Nordtvedt parameter indicating a violation of the
Strong Equivalence Principle in the PPN formalism.  The fact that
$\Psi$ is not equal to zero unless $\eta=0$ means that requirement
{\bf B} from Introduction is violated and the internal gravitational
field in our local reference system does not have the same form as the
corresponding solution for an isolated body.  This has a profound
physical meaning (see, also Section \ref{Nordtvedt-section} for
further discussion) and is directly related with the violation of the
equivalence principle for those theories where $\eta\neq0$.  In case
of a system consisting of two spherically symmetrical nonrotating
masses the term $\Psi$ is equal exactly to the 6th and 7th terms in
$g_{00}$ in Eq.  (52) in \cite{sah:ashby:88}, providing an
additional check of our results.

Note that the expression (\ref{Psi}) for $\Psi$ results from the
Newtonian potential $U_{\rm E}$ if we consider the expression
(\ref{G-local-linear}) for the ``effective'' gravitational constant as
seen in the local system.

\bigskip\noindent
2) Equating the external terms of order ${|\ve{X}|}^0$ give us the equation

\begin{eqnarray}\label{B}
&&{\dot B}(t)=-{1\over 8}v_E^4
+2(\gamma+1)\,v_E^i\,{\overline{w}}^i(\ve{x}_E)
-\left(\gamma+{1\over 2}\right)\,v_E^2\,{\overline{w}}(\ve{x}_E)
+\left(\beta-{1\over 2}\right)\,
{\overline{w}}^2(\ve{x}_E).
\end{eqnarray}

\bigskip\noindent
3) Matching the terms of order ${|\ve{X}|}^1$ leads to the equations
of motion of the origin of the local system with respect to the
barycentric PPN reference system

\begin{eqnarray}\label{aei}
&&a_E^i={\overline{w}}_{,i}(\ve{x}_E)
-R^a_{\ j}Q_a
\left(\delta^{ij}-{1\over c^2}\left(
v_E^2\,\delta^{ij}
+(2+\gamma)\,\overline{w}(\ve{x}_E)\,\delta^{ij}
+{1\over 2} v_E^i v_E^j
\right)\right)
\nonumber \\
&&
\phantom{a_E^i=}
+{1\over c^2}\biggl(
2(1+\gamma)\,\dot{{\overline{w}}}^i(\ve{x}_E)
+\left(\gamma v_E^2
-2(\gamma+\beta)\,{\overline{w}}(\ve{x}_E)\right)
\overline{w}_{,i}(\ve{x}_E)
-(2\gamma+1)\,v_E^i\,\dot{{\overline{w}}}(\ve{x}_E)
\nonumber \\
&&
\phantom{a_E^i(t^*)=+{1\over c^2}\biggl(}
-2(1+\gamma)\,v_E^j\,{\overline{w}}^j_{,i}(\ve{x}_E)
-v_E^i\,v_E^j\,\overline{w}_{,j}(\ve{x}_E)
\biggr)
+{\cal O}(c^{-4}).
\end{eqnarray}

\noindent
Note that the Newtonian terms ${\overline{w}}_{,i}(\ve{x}_E)$ and $Q_a$
in (\ref{aei}) should be computed at moments of time $t$ and $T$
related to each other by $T=t-1/c^2\,A(t)+\OO4$, that is by the
coordinate transformation (\ref{trans:time}) evaluated at
$\ve{x}=\ve{x}_E$.  For $Q_a=0$, (\ref{aei}) coincides with the PPN
equations of motion of a test particle (geodetic equation), provided
that the gravitational potentials of the local subsystem $w_{\rm E}$,
$w^i_{\rm E}$ are excluded from the metric.

We consider $Q_a$ as a known function of $T$ specified in the local
reference system $(T,X^a)$ with post-Newtonian accuracy. When computing
the Newtonian terms in (\ref{aei}) with post-Newtonian accuracy, $Q_a$
should be expressed in terms of global coordinates. It is easy to see
that all relativistic terms in (\ref{aei}) proportional to $Q_a$ are
due to the re-computation of $Q_a$ from the local reference system to
the global one.  Indeed, if a test particle is situated at the origin
of the local reference system ($X^a_{\rm test}=0$) and if its local
coordinate velocity is equal to zero ($dX^a_{\rm test}/dT=0$) then
making use of (\ref{trans:time})--(\ref{trans:space}) we get

\begin{eqnarray}\label{BRS:Q_i}
{d^2x^i_{\rm test}\over dt^2}-a_E^i&=&R^a_{\ j}\,{d^2X^a_{\rm
test}\over dT^2}
\,\left(\delta^{ij}-{1\over c^2}\left(v_E^2\,\delta^{ij}+(2+\gamma)\,
{\overline{w}}(\ve{x}_E)\,\delta^{ij}
+{1\over 2}\,v_E^i\,v_E^j\right)\right)
\nonumber\\
&&+\OO4,
\end{eqnarray}

\noindent
where $d^2x^i_{\rm test}/dt^2-a_E^i$ is the global coordinate
acceleration of the test particle relative to the origin of the local
reference system. The terms on the right-hand side of (\ref{BRS:Q_i})
exactly match those in (\ref{aei}) that are proportional to $Q_a$.
$Q_a$ is, thus, the acceleration of the instantaneous locally inertial
reference system (whose origin coincides with that of the local
reference system at a given moment of time) expressed in the local
reference system. This also can be seen from the equations of motion of
test particles relative to the local reference system (see, Section
\ref{section:eqm-test-particle} below). The equations of motion can be
represented as

\begin{equation}\label{ddot-w-Q}
\ddot X^a_{\rm test}=Q^a(T)+f^a(\ve{X}_{\rm test},\dot\ve{X}_{\rm test})+\OO4,
\end{equation}

\noindent
where $f^a(\ve{X}_{\rm test},\dot\ve{X}_{\rm test})$ is some force
depending on position and/or velocity of the particle (i.e., the only
term in the equations of motion independent on the actual trajectory of
the particle is $Q^a$). Note also that $Q_a$ can be directly related to
the 4-acceleration of the worldline of the origin of the local
reference system (see, e.g., \cite{dsx:91}).

\bigskip\noindent
4) Matching of the external terms of order ${|\ve{X}|}^l, l\geq2$
in closed form leads to an expression for the relativistic tidal
potential of the form

\begin{eqnarray}\label{W_T}
W_{\rm T}(T,\ve{X})=&&
{\overline{w}}(t,\ve{x})-{\overline{w}}(\ve{x}_E)
-{\overline{w}}_{,j}(\ve{x}_E)r_E^j
\nonumber \\
&&+{1\over c^2}\biggl(
-2(1+\gamma) v_E^i
\left({\overline{w}}^i(t,\ve{x})-{\overline{w}}^i(\ve{x}_E)
-{\overline{w}}^i_{,j}(\ve{x}_E)r_E^j \right)
+(1+\gamma) v_E^2 W_{\rm T}
\nonumber \\
&&
\phantom{+{1\over c^2}\biggl(}
+(1+\gamma)\dot{\overline{w}}^{i,j}(\ve{x}_E)\,r_E^i
\,r_E^j
+{1\over 2}\,\gamma\,\ddot{{\overline{w}}}(\ve{x}_E)\,r_E^2
+\left({1\over 2}-\beta-\gamma\right)\,{(a_E^i r_E^i)}^2
\nonumber \\
&&
\phantom{+{1\over c^2}\biggl(}
+(1-2\beta-2\gamma)\,Q_aX^a\,a_E^ir_E^i
-\gamma\,v_E^ir_E^i\,
\dot{\overline{w}}_{,j}(\ve{x}_E)\,r_E^j
+{1\over 2}\gamma\,r_E^2\,R^a_{\ i}a_E^i Q_a
\nonumber \\
&&
\phantom{+{1\over c^2}\biggl(}
+C_{,T}(T,\ve{X})
+2(1-\beta)\left({\overline{w}}(\ve{x}_E)
+a_E^i\,r_E^i\right) W_{\rm T}
\biggr)+{\cal O}(c^{-4}),
\\
\label{W_T^a}
W_{\rm T}^a(T,\ve{X})=&&R^a_{\ i}\,\biggl\{{\overline{w}}^i(t,\ve{x})-
{\overline{w}}^i(\ve{x}_E)
-{\overline{w}}^i_{,j}(\ve{x}_E)r_E^j
-v_E^i W_{\rm T}(T,\ve{X})
\nonumber \\
&&\phantom{R^a_{\ i}\,\biggl\{}
+{1\over 2(1+\gamma)}\left(
\gamma\,\left(r_E^i \dot a_E^j r_E^j
-{1\over 2}\,\dot a_E^i r_E^2\right)
-C_{,i}(t,\ve{x})\right)\biggr\}+{\cal O}(c^{-2}),
\end{eqnarray}

\noindent
where $C_{,T}$ is a partial derivative of $C$ with respect to the
local coordinate time $T$. E.g., for

\begin{equation}\label{C(T,X)}
C(T,\ve{X})=C_1(T,\ve{X})={\gamma-2\over 10}\ve{X}^2\,\dot a_E^i\,R^a_i\,X^a,
\end{equation}

\noindent
one gets

\begin{equation}\label{C,T}
C_{,T}(T,\ve{X})={\gamma-2\over 10}\ve{X}^2\,\ddot a_E^i\,R^a_i\,X^a.
\end{equation}

Let us note that in the Newtonian terms in the right-hand side
(\ref{W_T}) the moment of time $t$ depends on both $T$ and $\ve{X}$
according to (\ref{trans:time})--(\ref{trans:space}). This should be
accounted for when computing, e.g., partial derivatives of $W_{\rm T}$
with respect to $T$.

From (\ref{W_T}) it follows that

\begin{eqnarray} \label{WT,aa}
c^2\,W_{{\rm T},aa}&=&\ddot Q_a X^a+W_{{\rm T},TT}
-\eta\,\left(Q_a+W_{{\rm T},a}\right)\,R^a_{\ i}\,a_E^i
\nonumber \\
&&
+(\gamma-1)\,\left(\ddot{\overline w}(\ve{x}_E)
+R^a_{\ i}\,\ddot a_E^i\, X^a\right)
-2\,(\beta-1)\,a_E^2+\OO2,
\\ \label{WT,aabb}
W_{{\rm T},aabb}&=&\OO4.
\end{eqnarray}

\noindent
Eqs. (\ref{WT,aa}) and (\ref{WTa,bb}) can be considered as effective
field equations for the external potentials in local coordinates.

\bigskip
We have derived all unknown functions with post-Newtonian accuracy and
thereby obtained the local PPN metric tensor as well as
the transformations between the local and global PPN reference
systems.

\subsection{Energy-momentum tensor in the local reference system}

It is interesting to check the form of the energy-momentum tensor for
an ideal fluid (\ref{Tmunu}) in local coordinates, ${\cal
T}^{\alpha\beta}$. The energy-momentum tensor $T^{\alpha\beta}$ in the
global PPN reference system is related to ${\cal T}^{\alpha\beta}$ by

\begin{equation}\label{T-matching}
{\cal T}^{\alpha\beta}(T,\ve{X})=
{\partial X^\alpha\over \partial x^\mu}\,
{\partial X^\beta\over \partial x^\nu}\,
T^{\mu\nu}(t,\ve{x}).
\end{equation}

\noindent
For an ideal fluid $T^{\mu\nu}$ is defined by (\ref{Tmunu}) or
(\ref{Tmunu*}). Substituting the transformation rules
(\ref{trans:time})--(\ref{trans:space}) and the expressions (\ref{A}),
(\ref{D^{ij}}) and (\ref{D^{ijk}}) for $A$, $D^{ij}$ and $D^{ijk}$
respectively one gets

\begin{eqnarray}\label{Tmunu-GRS}
{\cal T}^{00}&=&\rho c^2\left[1+{1\over c^2}
\left(V^2+\Pi+2W\right)\right]+\OO2,
\nonumber \\
{\cal T}^{0a}&=&\rho c\, V^a\left[1+{1\over c^2}
\left(V^2+\Pi+2W\right)\right]+
{1\over c}\,p\,V^a+\OO3,
\nonumber \\
{\cal T}^{ab}&=&\rho\, V^a\, V^b+\delta^{ab} p+\OO2,
\end{eqnarray}

\noindent
where $V^a$ is the velocity of matter relative
to the local reference system, $\rho(T,\ve{X})=\rho(t,\ve{x})$ is the
invariant density, $\Pi(T,\ve{X})=\Pi(t,\ve{x})$ is the specific energy
density, and $p(T,\ve{X})=p(t,\ve{x})$ is the isotropic pressure,
$W(T,\ve{X})=w(t,\ve{x})-\overline{w}(\ve{x}_E)-a_E^i\,r_E^i+\OO2$.
One can also rewrite (\ref{Tmunu-GRS}) as

\begin{eqnarray}\label{Tmunu*-GRS}
{\cal T}^{00}&=&\hat\rho^* c^2\left[1+{1\over c^2}
\left({1\over 2}\,V^2+\Pi-(3\gamma-2) W\right)\right]+\OO2,
\nonumber \\
{\cal T}^{0a}&=&\hat\rho^* c\, V^a\left[1+{1\over c^2}\left(
{1\over 2}\,V^2+\Pi-(3\gamma-2) W\right)\right]+
{1\over c}\,p\,V^a+\OO3,
\nonumber \\
{\cal T}^{ab}&=&\hat\rho^*\,V^a\,V^b+\delta^{ab} p+\OO2,
\end{eqnarray}

\noindent
where

\begin{equation}\label{rho*-GRS}
\hat\rho^*(T,\ve{X})=\rho(T,\ve{X})\,
(-{\cal G})^{1/2}\,{c\,dT\over ds}=\rho\left[1+{1\over c^2}
\left({1\over 2}\,V^2+3\gamma W\right)+\OO4\right],
\end{equation}

\noindent
so that

\begin{equation}\label{hatrho*-cont}
{\partial \over \partial T} \hat\rho^*+{\partial \over \partial X^a}
\left( \hat\rho^*\,V^a \right)=\OO4.
\end{equation}

\noindent
Thus, the functional form of the energy momentum tensor is the
same in the global and local PPN reference systems.

\section{Other forms of the local metric}
\label{Nordtvedt-section}

In the previous Section we have succeeded to construct a local
reference system $(T,X^a)$ which satisfies property {\bf A} from
Section \ref{Introduction}, but does not satisfy property {\bf B}.
It is interesting to check if there exist other versions of a local
reference system satisfying properties {\bf A} and/or {\bf B}. To this
end let us consider another set of local coordinates $(\tilde T,\tilde
X^a)$ related to $(T,X^a)$ by the following transformations

\begin{eqnarray}\label{X-X'}
\tilde T&=&T,
\nonumber \\
\tilde X^a&=&X^a+{1\over c^2}\,{\eta\over\gamma}\,
\left(R^a_{\ i}\,R^b_{\ j}\,D^{ij} X^b
+R^a_{\ i}\,R^b_{\ j}\,R^c_{\ k}\,D^{ijk}
X^b X^c\right)+\OO4,
\end{eqnarray}

\noindent
where $D^{ij}$ and $D^{ijk}$ are defined by (\ref{D^{ij}})--(\ref{D^{ijk}}).
One can show that in $(\tilde T,\tilde X^a)$ the metric tensor reads

\begin{eqnarray}\label{GRS:metric:alternative}
\tilde G_{00}&=&-1+{2\over c^2}\,\tilde W(\tilde T,\tilde \ve{X})
-{2\over c^4}\,\beta\,\tilde W^2(\tilde T,\tilde \ve{X})+{\cal O}(c^{-5}),
\nonumber \\
\tilde G_{0a}&=&-{2(1+\gamma)\over c^3}\,
\tilde W^a(\tilde T,\tilde \ve{X})
-{1\over c^3}\,\eta\,
\left(\dot{\overline{w}}(\ve{x}_E)\,\tilde X^a
+\tilde X^a\,R^b_{\ i}\,\dot a_E^i\,\tilde X^b
-{1\over 2}\,R^a_{\ i}\,\dot a_E^i\,\tilde \ve{X}^2
\right)
+{\cal O}(c^{-5}),
\nonumber \\
\tilde G_{ab}&=&\delta_{ab}\left(1+{2\over c^2}\,\left(\gamma\,
\tilde W(\tilde T,\tilde\ve{X})-\eta \left(\overline{w}(\ve{x}_E)+
R^a_{\ i}\,a_E^i\,\tilde X^a\right)\right)\right)+{\cal O}(c^{-4}),
\end{eqnarray}

\begin{eqnarray}\label{tilde-GRS:split}\label{tilde-W-split}
\tilde W(\tilde T,\tilde \ve{X})&=&
\tilde W_{\rm E}(\tilde T,\tilde \ve{X})+\tilde Q_a(\tilde T)\,\tilde X^a+
\tilde W_{\rm T}(\tilde T,\tilde \ve{X}),
\\ \label{tilde-Wi-split}
\tilde W^a(\tilde T,\tilde \ve{X})&=&\tilde W^a_{\rm E}(\tilde T,\tilde \ve{X})
+{1\over 2}\,\varepsilon_{abc}\,\tilde C_b(\tilde T)\,\tilde X^c
+\tilde W^a_{\rm T}(\tilde T,\tilde \ve{X}),
\end{eqnarray}

\noindent
where

\begin{eqnarray}
\label{tilde-WE}
\tilde W_{\rm E}&=&G\,\int_V\,
\tilde \Sigma(\tilde T,\tilde \ve{X}')\,{1\over |\tilde \ve{X}-\tilde \ve{X}'|}\,
d^3\tilde X'
\nonumber\\
&&
+{1\over 2c^2}\,G\,{\partial^2\over\partial \tilde T^2}\,
\int_V\,\tilde \Sigma(\tilde T,\tilde \ve{X}')\,
|\tilde \ve{X}-\tilde \ve{X}'|\,d^3\tilde X'+{\cal O}(c^{-4}),
\\ \label{tilde-WEi}
\tilde W^a_{\rm E}&=&G\,\int_V\,\tilde \Sigma^a(\tilde T,\tilde \ve{X}')\,
{1\over |\tilde \ve{X}-\tilde \ve{X}'|}\,d^3\tilde X'
+{\cal O}(c^{-2}),
\end{eqnarray}

\begin{eqnarray}\label{tilde-Sigma}
\tilde \Sigma&=&
{1\over c^2}\left(\tilde {\cal T}^{00}+\gamma\,\tilde {\cal T}^{aa}+
{1\over c^2}\,\tilde {\cal T}^{00}\,(3\gamma-2\beta-1)\,\tilde W\right)+{\cal O}(c^{-4}),
\\ \label{tilde-Sigmai}
\tilde \Sigma^a&=&{1\over c}\,\tilde {\cal T}^{0a}+{\cal O}(c^{-2}),
\end{eqnarray}

\noindent
where $\tilde {\cal T}^{\mu\nu}$ are the components of the
energy-momentum tensor in coordinates $(\tilde T,\tilde\ve{X})$, and
$\tilde W_{\rm T}$ and $\tilde W^a_{\rm T}$ are some external tidal
($\sim {\cal O}(\tilde \ve{X}^2)$) potentials.

The internal potentials $\tilde W_E$ and $\tilde W^a_E$ have the same
form as the potentials of the corresponding isolated source provided
that the tidal influence of external metric is neglected and,
therefore, in coordinates $(\tilde T,\tilde\ve{X})$ property {\bf B}
from Section \ref{Introduction} is satisfied.  However, the terms
proportional to $\eta$ in $\tilde G_{0a}$ and $\tilde G_{ab}$
represent non-tidal external potentials, thus violating property
{\bf A}.

Although any reference system can be obviously used, we believe that
the original local coordinates $(T,\ve{X})$ are more convenient since a
violation of the Strong Equivalence Principle is presented solely by
the $\Psi$ term in $G_{00}$.  Moreover, the appearance of non-tidal
external potentials in $\tilde G_{0a}$ and $\tilde G_{ab}$ results in a
more complicated interpretation of observational data in $(\tilde
T,\tilde{\ve{X}})$ (e.g., involving a complicated redefinition of
locally measured physical constants and parameters).  In the following
we use a local reference system with coordinates $(T,\ve{X})$.

\section{Multipole expansions of the internal potentials}
\label{Section-multipoles-internal}

Although the metric tensor of the local PPN reference system described
above enables us to discuss any physical phenomena in this reference
system in the first post-Newtonian approximation, for many reasons an
expansion of the gravitational potentials in terms of multipole moments
is desirable.  The main usefulness of mass (and spin) multipole moments
of a body results from the fact that for astronomical, almost
'spheroidal' bodies usually only a few determine the equations of
motion to sufficient accuracy.  Numerical values of multipole moments
can be obtained directly from various astronomical observations without
evaluating volume integrals like (\ref{WE}) appearing in the local
metric tensor.

The complete solution of the problem of multipole expansions of
gravitational potentials and equations of motion has been found in
\cite{dsx:91,dsx:92,dsx:93} for the case of General Relativity.  The
authors use the so called Blanchet-Damour (BD) mass multipole moments
\cite{blan:damour:86,blan:damour:89}. The BD moments are defined
physically by reference to multipole moments which can be read off
from gravitational waves emitted by an isolated system of matter. The
principal attractive property of the BD moments for their practical
applications in multipole expansions is contained in Theorem 6 of
\cite{dsx:91}. The Theorem states that the multipole expansions of
general-relativistic post-Newtonian gravitational potentials in terms
of the BD multipole moments are almost of Newtonian form
\cite{dsx:91}. In \cite{dsx:91,dsx:92,dsx:93} it has been shown that
this almost Newtonian form of multipole expansions is advantageous for
both translational and rotational equations of motion.

It turns out that in the PPN formalism we can find adequate multipole
moments by ``mimicking'' the almost Newtonian form of multipole
expansion of the post-Newtonian gravitational potential. It is easy to see
that with the following definition of generalized BD moments

\begin{eqnarray}\label{ML-BD-PPN}
{\cal M}_L&=&\int_V \Sigma\, \hat X^L d^3X
+{1\over 2(2l+3)} {1\over c^2}\, {d^2\over dT^2}
\int_V \Sigma\, \hat X^L  X^2 d^3X
\nonumber \\
&&
-{2(\gamma+1)(2l+1)\over (l+1)(2l+3)} {1\over c^2}\,
{d\over dT} \int_V \Sigma^a \hat X^{aL} d^3X,\quad l \ge 0,
\\ \label{SL-BD-PPN}
{\cal S}_L&=&\int_V \varepsilon^{bc\langle a_l} \hat X^{L-1\rangle b}
\,\Sigma^c\,d^3X,\quad l \ge 1,
\end{eqnarray}

\noindent
one gets

\begin{eqnarray}\label{BD-expansion-PPN}
W_E&=&G\sum_{l=0}^\infty {(-1)^l\over l!}
\left[ {\cal M}_L\,\partial_L {1\over |\ve{X}|}+
{1\over 2c^2}\,\ddot {\cal M}_L
\,\partial_L |\ve{X}|\right]
+{2(1+\gamma)\over c^2}\Lambda_{,T}+\OO4,
\end{eqnarray}

\begin{eqnarray}\label{BD-expansion-PPN-Wi}
2(1+\gamma) W^a_E&=&-2(1+\gamma)G\sum_{l=1}^\infty {(-1)^l\over l!}
\left[
\dot{\cal M}_{aL-1} \partial_{L-1} {1\over |\ve{X}|}+
{l\over l+1} \varepsilon_{abc} {\cal S}_{cL-1}
\partial_{bL-1} {1\over |\ve{X}|}
\right]
\nonumber \\
&&-2(1+\gamma)\Lambda_{,a}+\OO2,
\end{eqnarray}

\noindent
where

\begin{eqnarray}\label{Lambda}
\Lambda&=&G \sum_{l=0}^\infty
{(-1)^l\over (l+1)!}\,{2l+1\over 2l+3}
\,{\cal P}_L\,\partial_L {1\over |\ve{X}|},
\end{eqnarray}

\begin{eqnarray}\label{PL}
{\cal P}_L=\int_V \Sigma^a \, \hat X^{aL}\,d^3X.
\end{eqnarray}

\noindent
The terms containing $\Lambda$ can be eliminated from
(\ref{BD-expansion-PPN})--(\ref{BD-expansion-PPN-Wi})
by a transformation of time coordinate

\begin{equation}\label{kill-Lambda-PPN}
T'=T-{2(1+\gamma)\over c^4} \Lambda.
\end{equation}

\noindent
This gauge is called skeletonized harmonic gauge in \cite{dsx:91} (no
$\Lambda$-terms appear in
(\ref{BD-expansion-PPN})--(\ref{BD-expansion-PPN-Wi}) and the
gravitational potentials $W_{\rm E}$ and $W_{\rm E}^a$ are
``skeletonized'' by the moments ${\cal M}_L$ and ${\cal S}_L$).  The
generalized BD multipole moments (\ref{ML-BD-PPN})--(\ref{SL-BD-PPN})
obviously do not possess all nice properties of the BD moments in
General Relativity.  However, the expansion (\ref{BD-expansion-PPN})
has the same almost Newtonian form as the corresponding expansion in
General Relativity.  This fosters the hope that the generalized BD
multipole moments are suitable for the use in multipole expanded PPN
equations of motion.

\section{Tidal expansions of the external potentials}
\label{Section-multipoles-external}

Although we have derived the local external potentials $W_{\rm T}$ and
$W^i_{\rm T}$ in closed form, in many cases it is preferable to have
tidal expansions of them. These expansions are very
convenient, e.g., for a discussion of equations of motions in
the local reference system.  Tidal expansions of the external
potentials entering $G_{00}$ and $G_{0i}$ are quite straightforward to
derive and read

\begin{eqnarray}\label{G00:STF expansion}
Q_a\,X^a+W_{\rm T}&=&
\sum_{l=1}^\infty {1\over l!} Q_L\,\hat X^L
-{1\over c^2}
\sum_{l=3}^\infty {1\over l!}\,\left(\dot R_L-\dot V_L\right)\,\hat X^L
+{1\over 2c^2}
\sum_{l=0}^\infty {1\over l!}\,{X^2\over 2l+3}
K_L\hat X^L,
\end{eqnarray}

\begin{eqnarray}\label{G0i:STF expansion}
2\,(1+\gamma)\,W^a_{\rm T}&=&
2\,(1+\gamma)\,\sum_{l=2}^\infty {l\over (l+1)!}
\varepsilon_{abc_l}\,C_{bL-1}\,\hat X^L
\nonumber \\
&&-2\,(1+\gamma)\,\sum_{l=1}^\infty {1\over (l+1)!}\,{2l+1\over 2l+3}
\,\dot Q_L \hat X^{aL}
\nonumber \\
&&+\sum_{l=2}^\infty {1\over l!}\,\left(R_{aL}-V_{aL}\right)\,\hat X^{L}
\nonumber \\
&&
+{1\over 5}\,(\gamma-1)\,\left(R^b_{\ i}\,\dot a_E^i\,X^a\,X^b
-2 R^a_{\ i}\,\dot a_E^i X^2\right)+\OO2,
\end{eqnarray}

\noindent
where the symmetric tracefree tidal moments $Q_L$, $C_L$, $K_L$ and
$R_L$ are defined as

\begin{eqnarray}\label{Q_L-pN}
Q_L&=&{\rm STF}_L\,\biggl\{d_L\,\overline{w}(\ve{x}_E)
+{1\over c^2}\,\biggl(
2(1+\gamma)\,R^{a_l}_{\ i}\, d_{L-1}\,\dot{\overline{w}}^i(\ve{x}_E)
-2\,(1+\gamma)\,v_E^i\,
d_L\,\overline{w}^i(\ve{x}_E)
\nonumber \\
&&\phantom{{\rm STF}_L\,\biggl\{}
+\,(1+\gamma)\,v_E^2\,d_L\,\overline{w}(\ve{x}_E)
-{1\over 2}\,l\,R^{a_{l}}_{\ i}\,R^b_{\ j}\,v_E^i\,v_E^{j}\,
d_{bL-1}\,\overline{w}(\ve{x}_E)
+\lambda^L
\nonumber \\
&&\phantom{{\rm STF}_L\,\biggl\{}
-\Bigl[\gamma\,l+2\,(\beta-1)\Bigr]\,\overline{w}(\ve{x}_E)\,
d_L\,\overline{w}(\ve{x}_E)
+\Bigl[l-2\,(1+\gamma)\Bigr]\,R^{a_l}_{\ i}\,v_E^i\,
d_{L-1}\,\dot{\overline{w}}(\ve{x}_E)
\nonumber \\
&&\phantom{{\rm STF}_L\,\biggl\{}
 -\Bigl[\gamma\,l\,(l-1)+2\,(\beta-1)\,l+2\,(1+\gamma)\Bigr]
\,R^{a_l}_{\ i}\,a_E^{i}\,
d_{L-1}\,\overline{w}(\ve{x}_E)
\biggr)\biggr\}
\nonumber \\
&&\phantom{{\rm STF}_L\,\biggl\{}
+\OO4,\quad l\ge 2,
\end{eqnarray}

\begin{eqnarray}\label{lambda}
\lambda^{ab}&=&
(2\beta+2\gamma-1)\,R^a_{\ i}\,R^b_{\ j}\,a_E^{i}\,a_E^{j},
\nonumber \\
\lambda^{L}&=&0, \quad l\ge 3,
\end{eqnarray}

\begin{eqnarray}\label{R_L}
R_{L}&=&2\,(1+\gamma)\,{\rm STF}_L\left\{
\,R^{a_1}_{\ i}
\left[d_{L-1}\,{\overline{w}}^i(\ve{x}_E)
- v_E^i\,d_{L-1}\,{\overline{w}}(\ve{x}_E)
\right]\right\}+\OO2, \qquad l\geq 3,
\end{eqnarray}

\begin{eqnarray}\label{K}
K&=&(\gamma-1)\ddot{\overline U}(\ve{x}_E)+2\,(1-\beta)\,a_E^2
-\eta\,Q_a\,R^a_{\ i}\,a_E^i+\OO2,
\\ \label{Ki}
K_a&=&\ddot Q_a-\eta\,Q_{ab}\,R^b_{\ i}\,a_E^i+(\gamma-1)\,R^a_{\ i}
\,\ddot a_E^i+\OO2,
\\ \label{K_L,l>2}
K_{L}&=&\ddot Q_{L}-\eta\,Q_{aL}\,R^a_{\ i}\,a_E^i+\OO2,
\quad l\ge2,
\end{eqnarray}

\begin{eqnarray}\label{C_L}
C_L&=&{\rm STF}_L\,\left\{
\varepsilon_{a_lbc}\,R^c_{\ i}\,d_{bL-1}\,
\left({\overline{w}}^i(\ve{x}_E)-v_E^i\,\overline{w}(\ve{x}_E)\right)
\right\},\ l\ge2,
\end{eqnarray}

\noindent
(note that the expression in braces is already trace free).
Here $d_L$ is the shorthand notation

\begin{eqnarray}\label{d_L}
d_L\,\Phi(t,\ve{x})\equiv d_{a_1\dots a_l}\,\Phi(t,\ve{x})
\equiv R^{a_1}_{\ i_1}\dots R^{a_l}_{\ i_l}\,
{\partial^l \over \partial x^{i_1}\dots\partial x^{i_l}}\,\Phi(t,\ve{x}),
\end{eqnarray}

\noindent
and $d_L \Phi(\ve{x}_E)$ should be understood as the corresponding
derivative evaluated at $\ve{x}=\ve{x}_E$. In order to compute $Q_L$
for some given moment of local time $T$ with the aid of (\ref{Q_L-pN})
one has to evaluate $d_L\,\overline{w}(\ve{x}_E)$ at moment of global
time $t$ related with $T$ as $T=t-1/c^2\, A(t)+\OO4$, that is by the
transformations (\ref{trans:time}) evaluated at $\ve{x}=\ve{x}_E$.

Expansion (\ref{G00:STF expansion}) with (\ref{Q_L-pN})--(\ref{K})
can be derived by differentiating (\ref{W_T}) and using
(\ref{WT,aa})--(\ref{WT,aabb}) and the fact that for any $S$

\begin{eqnarray}\label{WT+Olaplace2}
S&=&\sum_{l=1}^\infty\,{1\over l!}\, \left.{\partial^l\over
\partial X^{a_1}\dots X^{a_l}}\, S\right|_{\ve{X}=0}\,\hat X^L
\nonumber\\
&&+{1\over l!}\,{1\over 2(2l+3)}\, \left.{\partial^l\over \partial
X^{a_1}\dots X^{a_l}}\, S_{,bb}\right|_{\ve{X}=0} \,\hat X^L\,\ve{X}^2+
{\cal O}(S_{,bbcc}).
\end{eqnarray}

\noindent
The moments $K_L$ represent the multipole expansion of (\ref{WT,aa}):
$c^{-2}\,K_L=W_{{\rm T},aaL}$. The quantities $V_L$ are arbitrary STF moments
appearing in the multipole expansion (\ref{C:general})--(\ref{C_2}) of
the function $C(T,\ve{X})$. One can get rid of the moments $R_L$ in
(\ref{G00:STF expansion})--(\ref{G0i:STF expansion}) by setting
$V_L=R_L$, that is, by choosing a special solution for $C(T,\ve{X})$

\begin{equation}\label{C:Br&Kop}
C(T,\ve{X})={\gamma-2\over 10} X^2 R^a_{\ i}\,\dot a_E^i\,X^a
+\sum_{l=3}^\infty {1\over l!} R_L\,\hat X^L.
\end{equation}

In the limit of General Relativity $\beta=\gamma=1$ only the two
families $Q_L$ and $C_L$ play a role in (\ref{G00:STF
expansion})--(\ref{G0i:STF expansion}).  These are related to the tidal
moments $G_L$ and $H_L$ of the DSX formalism \cite{dsx:91} as

\begin{eqnarray}\label{GH-QC}
G_L&=&Q_L\,\biggr|_{\beta=\gamma=1}\,,
\\
\nonumber \\
H_L&=&-4\,C_L\,\biggr|_{\beta=\gamma=1}\,.
\end{eqnarray}

\noindent
Because of the gauge dependence the family $R_L$ does not appear, e.g.,
in the equations of motion of test particles and $N$ extended bodies
which we derive in the next Sections. This can be shown also by computing
the tidal expansions of the following quantities which directly enter
the post-Newtonian equations of motion and are independent of the gauge
of the time coordinate

\begin{eqnarray}\label{gauge-inv-E}
&Q^a+&W_{{\rm T},a}+{1\over c^2}
\,2\,(1+\gamma)\,W^a_{{\rm T},T}=
\nonumber\\
&&
\sum_{l=0}^{\infty} {1\over l!}\, Q_{aL} \hat X^L
+{1\over c^2}\,2\,(1+\gamma)\,
\sum_{l=2}^\infty{1\over l!}\,{l\over l+1}\,
\varepsilon_{abc}\,\dot C_{bL-1}\,\hat X^{cL-1}
\nonumber\\
&&
+{1\over c^2}\,\sum_{l=0}^\infty{1\over l!}\,{1\over 2(2l+3)}\,
\ddot Q_{aL}\, \ve{X}^2\,\hat X^L
-{1\over c^2}\,\sum_{l=2}^\infty{1\over l!}\,
{(4\gamma+3)\,l-2\,(1+\gamma)\over 2l+1}\,
\ddot Q_{L-1}\,\hat X^{aL-1}
\nonumber\\
&&
+{1\over 3c^2}\,X^a
\left(\,(\gamma-1)\,\ddot{\overline w}(\ve{x}_E)+2\,(1-\beta)\,a_E^2\right)
-{1\over 6c^2}\,(\gamma-1)\,R^a_{\ i}\,\ddot a_E^i \ve{X}^2
\nonumber \\
&&
+{2\over 5c^2}\,(\gamma-1)\,R^b_{\ i}\,\ddot a_E^i\,\hat X^{ab}
-{1\over c^2}\,\eta\sum_{l=0}^\infty{1\over l!}\,{1\over 2l+3}\,
R^b_{\ i}\,a_E^i\,Q_{bL}\,\hat X^{aL}
\nonumber \\
&&
-{1\over 2c^2}\,\eta\sum_{l=0}^\infty {1\over l!}\,
{1\over 2l+3}\,R^b_{\ i}\,a_E^i\,Q_{abL}\,\ve{X}^2\,\hat X^L+\OO4,
\end{eqnarray}

\begin{eqnarray}\label{gauge-inv-B}
-2\,(1+\gamma)\,\varepsilon_{abc}\,W^b_{{\rm T},c}&=&
2\,(1+\gamma)\sum_{l=1}^\infty{1\over l!}\,\left( C_{aL}\, \hat X^L+
{l\over l+1}\varepsilon_{abc}\,\hat X^{bL-1}\,\dot Q_{cL-1}\right)
\nonumber\\
&&
+(\gamma-1)\,\varepsilon_{abc}\,R^b_{\ i}\,\dot a_E^i\,X^c
+\OO2.
\end{eqnarray}

\noindent
In the limit of General Relativity (\ref{gauge-inv-E}) and
(\ref{gauge-inv-B}) coincide with Equation (6.23a) and (6.23b) of
\cite{dsx:91}, respectively.

\section{Translational equations of motion of a test particle
in the local reference system}
\label{section:eqm-test-particle}

\subsection{The equations of motion in closed form}

  It is well know that a test particle moves along a geodetic
of spacetime as determined by the metric tensor. The geodetic equation
in the metric (\ref{GRS:metric}) reads

\begin{eqnarray}\label{eqm:W&Wa}
\ddot{X}^a_{\rm s}&=W_{,a}+
{\displaystyle {1\over c^2}}
\Bigl\{&
2(1+\gamma)\,W^a_{,T}
-2(\beta+\gamma)\,W\,W_{,a}
-(2\gamma+1)\,W_{,T}\,\dot{X}^a_{\rm s}
-2(1+\gamma) \,W_{,b}\,\dot{X}^b_{\rm s}\dot{X}^a_{\rm s}
\nonumber \\
&&
+\gamma\,W_{,a}\,\dot{X}^b_{\rm s}\,\dot{X}^b_{\rm s}+
2(1+\gamma)\,\left(W^a_{,b}-W^b_{,a}\right)\dot{X}^b_{\rm s}\Bigr\}
+\OO4.
\end{eqnarray}

\noindent
Here, $\ve{X}_{\rm s}$ are local coordinates of the test particle
(e.g., a satellite), and all potentials and their derivatives should be
evaluated at $(T,\ve{X})=(T,\ve{X}_{\rm s}(T))$. Accounting for the
split (\ref{W-split})--(\ref{Wi-split}) of the local potentials
entering into (\ref{GRS:metric}) it can be seen that the equations of
motion in the local reference system read

\begin{eqnarray}\label{ppN-eqm-struct}
\ddot{X}^a_{\rm s}&=&\Phi^a_{\rm E}+\Phi^a_{\rm el}
+{1\over c^2}\left(
\Phi^a_{\rm coup}+
\Phi^a_{\rm mg}+
\Phi^a_{\Psi}
\right)+\OO4,
\end{eqnarray}

\noindent
where $\Phi^a_{\rm E}$ are the terms describing the Earth's
gravitational field in the absence of an external world

\begin{eqnarray}\label{Phi-E}
\Phi^a_{\rm E}&=W_{{\rm E},a}+\displaystyle{1\over c^2}\biggl(
&2\,(1+\gamma)\,W^a_{{\rm E},T}
-2\,(\gamma+\beta)\,W_{\rm E}\,W_{{\rm E},a}
\nonumber \\
&&-(2\gamma+1)\,W_{{\rm E},T}\,\dot{X}^a_{\rm s}
-2\,(1+\gamma)\,W_{{\rm E},b}\,\dot{X}^b_{\rm s}\,\dot{X}^a_{\rm s}
+\,\gamma\,W_{{\rm E},a}\,\dot{X}^b_{\rm s}\,\dot{X}^b_{\rm s}
\nonumber\\
&&+\,2\,(1+\gamma)\,\dot{X}^b_{\rm s}\,
\left(W^a_{{\rm E},b}-W^b_{{\rm E},a}\right)
\biggr)+\OO4,
\end{eqnarray}

\noindent
The Earth-third body coupling term $\Phi^a_{\rm coup}$ due to the only
nonlinear term in (\ref{eqm:W&Wa}) reads

\begin{eqnarray}\label{Phi-coup}
\Phi^a_{\rm coup}=-2\,(\gamma+\beta)\,\biggl(W_{\rm E}\,
\left(Q_a+W_{{\rm T},a}\right)
+W_{{\rm E},a}\,\left(Q_a\,X^a_{\rm s}+W_{\rm T}\right)\biggr)+\OO2.
\end{eqnarray}

\noindent
The ``gravito-electric'' part $\Phi^a_{\rm el}$ of the influence of
external bodies (independent of the velocity of the satellite)
reads

\begin{eqnarray}\label{eqm:el}
\Phi^a_{\rm el}=Q_a+W_{{\rm T},a}
+{1\over c^2}\,\biggl(&&
2(1+\gamma)\,W_{{\rm T},T}^a
-2(\beta+\gamma)\,\left(Q_b\,X^b_{\rm s}+W_{\rm T}\right)\,
(Q_a+W_{{\rm T},a})
\nonumber \\
&&
+(1+\gamma)\,\varepsilon_{abc}\,\dot C_b\,X^c_{\rm s}
\biggr)+\OO4,
\end{eqnarray}

\noindent
and the ``gravito-magnetic'' part $\Phi^a_{\rm mg}$ (depending on the
velocity of the satellite) of the influence of external
bodies is given by

\begin{eqnarray}\label{eqm:mg}
\Phi^a_{\rm mg}&=&
-\,(1+2\gamma)\,\left(W_{{\rm T},T}+\dot Q_b\,X^b_{\rm s}\right)\,
\dot{X}^a_{\rm s}
-2\,(1+\gamma)\,\left(W_{{\rm T},b}+Q_b\right)\,\dot{X}^b_{\rm s}\,
\dot{X}^a_{\rm s}
\nonumber\\
&&+\,\gamma\,\left(W_{{\rm T},a}+Q_a\right)\,\dot{X}^b_{\rm s}\,
\dot{X}^b_{\rm s}
+\,4(1+\gamma)\,W_{\rm T}^{[a,b]}\,\dot{X}^b_{\rm s}
+2(1+\gamma)\,\varepsilon_{abc}\,C_b\,\dot X^c_{\rm s}
+\OO2,
\end{eqnarray}

\noindent
Let us remind that $\Phi^a_{\rm el}$ and $\Phi^a_{\rm mg}$ contain
also the two arbitrary functions $Q_a$ and $C_a$ that characterize the
world line of the local reference system and rotation of its spatial
axes.  A reasonable choice of $Q_a$ leading to vanishing of the BD
mass dipole ${\cal M}^a$ of the central body will be discussed in
Section \ref{LEQM-expanded}.  As it was noted above (see the
discussion after (\ref{Rij-Ca})) a dynamically nonrotating local
reference system corresponds to $C_a=0$ while in a kinematically
nonrotating local reference system $C_a$ contains the geodetic,
Lense-Thirring and Thomas precessions.

The term $\Phi^a_{\Psi}$ represents the additional acceleration due
to the term (\ref{Psi})

\begin{eqnarray}\label{Phi-Psi}
\Phi^a_{\Psi}(T,\ve{X})&=&\Psi_{,a}(T,\ve{X})
\nonumber\\
&=&-\eta\,\left(W_{{\rm E},a}
\left({\overline{w}}(\ve{x}_E)+ a_E^i\,r_E^i\right)
-\chi^{\rm E}_{,ab}(T,\ve{X})\,R^b_{\ i}\,a_E^i
+W_{\rm E}\,R^a_{\ i}\,a_E^i\right)+\OO2.
\end{eqnarray}

\noindent
In case of two spherically symmetrical nonrotating masses the term
$\Psi$ (the potential of $\Phi^a_{\Psi}$) coincides with the 6th and
7th terms in $g_{00}$ of Eq. (52) in \cite{sah:ashby:88}. A discussion
of observability of this effect in satellite motion data can be found
in Section VI of \cite{sah:ashby:88}.

\subsection{Multipole expansions of the equations of motion}

For many applications it is convenient to expand the equations of
motion derived in the previous Section into multipole moment of the
gravitational field of the central body and into tidal moments of the
external potentials. E.g., it is a common practice to use the equations
of motion of near-Earth satellites in expanded form and to fit the
multipole moments of the Earth's gravitational field.  Substituting the
expansions derived in Sections \ref{Section-multipoles-internal} and
\ref{Section-multipoles-external} into (\ref{Phi-E})--(\ref{eqm:mg})
and (\ref{Phi-Psi}) one gets

\begin{eqnarray}\label{Phi-E-mult}
\Phi^a_{\rm E}&=&
G\sum_{l=0}^{\infty} {(-1)^l\over l!}\, {\cal M}_L\,
\partial_{aL}\,{1\over |\ve{X}_{\rm s}|}
\nonumber\\
&&
+{1\over c^2}\,G\, \sum_{l=0}^{\infty} {(-1)^l\over l!}\,
\Biggl(
{1\over 2}\,\ddot {\cal M}_L\,\partial_{aL}\,|\ve{X}_{\rm s}|
+{2(1+\gamma)\over l+1}\, \ddot {\cal M}_{aL}\,
\partial_{L}\,{1\over |\ve{X}_{\rm s}|}
-{2(1+\gamma)\over l+2}\,\varepsilon_{abc}\,\dot {\cal S}_{bL}\,
\partial_{cL}\,{1\over |\ve{X}|_{\rm s}}
\nonumber\\
&&
\phantom{+{1\over c^2}\,G\,\sum_{l=0}^{\infty} {(-1)^l\over l!}\,\Biggl(}
-2(\gamma+\beta)\,{\cal M}_L\,\partial_{aL}\,{1\over |\ve{X}_{\rm s}|}\,
\left(G \sum_{k=0}^\infty {(-1)^k\over k!}\, {\cal M}_K\,
\partial_{K}\,{1\over |\ve{X}_{\rm s}|}\right)
\nonumber\\
&&
\phantom{+{1\over c^2}\,G\,\sum_{l=0}^{\infty} {(-1)^l\over l!}\,\Biggl(}
+\gamma\,\dot X^b_{\rm s}\,\dot X^b_{\rm s}\,{\cal M}_L\,
\partial_{aL}\,{1\over |\ve{X}_{\rm s}|}\,
-2(1+\gamma)\,\dot X^b_{\rm s}\,\dot X^a_{\rm s}\,{\cal M}_L\,
\partial_{bL}\,{1\over |\ve{X}_{\rm s}|}\,
\nonumber\\
&&
\phantom{+{1\over c^2}\,G\,\sum_{l=0}^{\infty} {(-1)^l\over l!}\,\Biggl(}
-(2\gamma+1)\,\dot X^a_{\rm s}\,\dot{\cal M}_L\,
\partial_{L}\,{1\over |\ve{X}_{\rm s}|}
+{4(1+\gamma)\over l+1}\,\dot X^b_{\rm s}\,\dot {\cal M}_{L[a}\,
\partial_{b]L}\,{1\over |\ve{X}_{\rm s}|}
\nonumber\\
&&
\phantom{+{1\over c^2}\,G\,\sum_{l=0}^{\infty} {(-1)^l\over l!}\,\Biggl(}
+{4(1+\gamma)\over l+2}\,\dot X^b_{\rm s}\,{\cal S}_{cL}\,\varepsilon_{dc[a}
\,\partial_{b]dL}\,{1\over |\ve{X}_{\rm s}|}
\Biggr)
+\OO4,
\end{eqnarray}

\begin{eqnarray}\label{Phi-coup-mult}
\Phi^a_{\rm coup}&=-2\,(\gamma+\beta)\,G\,\Biggl(&
\sum_{l=0}^\infty\sum_{k=0}^\infty {(-1)^l\over l!}\,{1\over k!}\,
{\cal M}_L\,Q_{aK}\, \hat X^K_{\rm s}\,\partial_L\,{1\over |\ve{X}_{\rm s}|}
\nonumber\\
&&
+\sum_{l=0}^\infty\sum_{k=1}^\infty {(-1)^l\over l!}\,{1\over k!}\,
{\cal M}_L\,Q_{K}\,\hat X^K_{\rm s}\,\partial_{aL}\,{1\over |\ve{X}_{\rm s}|}
\Biggr)+\OO2,
\end{eqnarray}

\begin{eqnarray}\label{eqm:el-mult}
\Phi^a_{\rm el}&=&
\sum_{l=0}^{\infty} {1\over l!}\, Q_{aL} \hat X^L_{\rm s}
+{1\over c^2}\,\Biggl(
2\,(1+\gamma)\,\sum_{l=1}^\infty{1\over l!}\,{l\over l+1}\,
\varepsilon_{abc}\,\dot C_{bL-1}\,\hat X^{cL-1}_{\rm s}
\nonumber\\
&&
+\sum_{l=0}^\infty{1\over l!}\,{1\over 2(2l+3)}\,
\ddot Q_{aL}\, \ve{X}_{\rm s}^2\,\hat X^L_{\rm s}
-\sum_{l=2}^\infty{1\over l!}\,
{(4\gamma+3)\,l-2\,(1+\gamma)\over 2l+1}\,
\ddot Q_{L-1}\,\hat X^{aL-1}_{\rm s}
\nonumber\\
&&
+{1\over 3}\,X^a_{\rm s}
\left(\,(\gamma-1)\,\ddot{\overline U}(\ve{x}_E)+2\,(1-\beta)\,a_E^2\right)
-{1\over 6}\,(\gamma-1)\,R^a_{\ i}\,\ddot a_E^i\, \ve{X}^2_{\rm s}
\nonumber \\
&&
+{2\over 5}\,(\gamma-1)\,R^b_{\ i}\,\ddot a_E^i\,\hat X^{ab}_{\rm s}
-\eta\,\sum_{l=0}^\infty{1\over l!}\,{1\over 2l+3}\,
R^b_{\ i}\,a_E^i\,Q_{bL}\,\hat X^{aL}_{\rm s}
\nonumber \\
&&
-{1\over 2}\,\eta\,\sum_{l=0}^\infty {1\over l!}\,
{1\over 2l+3}\,R^b_{\ i}\,a_E^i\,Q_{abL}\,\ve{X}_{\rm s}^2\,\hat X^{L}_{\rm s}
\nonumber \\
&&
-2(\gamma+\beta)\,
\sum_{l=0}^\infty\,\sum_{k=1}^\infty
{1\over l!}\,{1\over k!}\,Q_{aL}\,Q_K\,\hat X^L_{\rm s}\,\hat X^K_{\rm s}
\Biggr)
+\OO4,
\end{eqnarray}

\begin{eqnarray}\label{eqm:mg-mult}
\Phi^a_{\rm mg}&=&
-(2\gamma+1)\,\dot X^a_{\rm s}\,
\sum_{l=1}^\infty {1\over l!}\,\dot Q_L\,\hat X^L_{\rm s}
-2(1+\gamma)\,\dot X^a_{\rm s}\,\dot X^b_{\rm s}\,\sum_{l=0}^\infty
{1\over l!}\,Q_{bL}\,\hat X^L_{\rm s}
+\gamma\,\dot X^b_{\rm s}\,\dot X^b_{\rm s}\,
\sum_{l=0}^\infty {1\over l!}\,Q_{aL}\,\hat X^L_{\rm s}
\nonumber\\
&&
-2(1+\gamma)\,\varepsilon_{abc}\,\dot X^b_{\rm s}\,
\sum_{l=0}^\infty {1\over l!}\,
\left(C_{cL}\,\hat X^L_{\rm s}
+{1\over l+2}\,\varepsilon_{cde}\,\hat X^{dL}_{\rm s}\,\dot Q_{eL}\right)
\nonumber\\
&&
+(\gamma-1)\,(X^a_{\rm s}\,R^b_{\ i}\,\dot a^i_E\,\dot X^b_{\rm s} -
R^a_{\ i}\,\dot a^i_E\,X^b_{\rm s}\,\dot X^b_{\rm s})
+\OO2,
\end{eqnarray}

\noindent
where we use (\ref{gauge-inv-E})-(\ref{gauge-inv-B}) and the identity
$2\,W_{\rm T}^{[a,b]}=\varepsilon_{abc}\,\varepsilon_{cde}\,W_{\rm T}^{d,e}$.

The multipole expansion of $\Phi^a_{\Psi}$ reads

\begin{eqnarray}\label{Phi-Psi-mult}
\Phi^a_{\Psi}=&-\eta\,G\,\displaystyle{\sum_{l=0}^\infty {(-1)^l\over l!}}
\biggl\{\,&{\cal M}_L\,\biggl(
\left(\overline w(\ve{x}_E)+R^b_{\ i}\,a_E^i\,X^b_{\rm s}\right)\,
\partial_{aL}\,{1\over |\ve{X}_{\rm s}|}
\nonumber\\
&&\phantom{{\cal M}_L\,\biggl(}
+R^a_{\ i}\,a_E^i\,\partial_L\,{1\over |\ve{X}_{\rm s}|}
-{1\over 2}\,R^b_{\ i}\,a_E^i\,\partial_{abL}\,|\ve{X}_{\rm s}|
\biggr)
\nonumber \\
&&-{1\over 2(2l+3)}\,{\cal N}_L\,R^b_{\ i}\,a_E^i\,
\partial_{abL}\,{1\over |\ve{X}_{\rm s}|}
\biggr\}
+\OO2,
\end{eqnarray}

\noindent
where

\begin{eqnarray}\label{NL}
{\cal N}_L=\int_V \Sigma\, \ve{X}^2\,\hat X^L\, d^3X.
\end{eqnarray}

\noindent
Here we use the multipole expansion of $\chi^{\rm E}(T,\ve{X})$

\begin{equation}\label{chi-mult}
\chi^{\rm E}(T,\ve{X})={1\over 2}\,G\sum_{l=0}^\infty
{(-1)^l\over l!}\,\left({\cal M}_L\,\partial_L |\ve{X}|+
{1\over 2l+3}\,{\cal N}_L\,\partial_L\,{1\over |\ve{X}|}\right)+\OO2,
\end{equation}

\noindent
which can be verified directly from (\ref{chiE-geo}).
The following relations are useful for evaluation of the
expansions derived above

\begin{eqnarray}\label{partial-L-1-over-X}
\partial_L\,{1\over |\ve{X}|}&=&(-1)^l\,(2l-1)!!\,{\hat N^L\over
|\ve{X}|^{l+1}},
\\ \label{partial-L-X}
\partial_{L}\,|\ve{X}|&=&(-1)^{l-1}\,(2l-3)!!\,{1\over |\ve{X}|^{l-1}}\,
\left(\hat N^L-{l(l-1)\over(2l-1)(2l-3)}\,\delta^{(a_{l-1}a_l}\hat N^{L-2)}
\right),
\end{eqnarray}

\noindent
where $N^a=X^a/|\ve{X}|$ and $\hat N^L=N^{\langle a_1}
N^{a_2}\dots\,N^{a_l\rangle}$. The shorthand notations $\partial_L
(1/|\ve{X}_{\rm s}|)$ and $\partial_L |\ve{X}_{\rm s}|$ mean
$\partial_L (1/|\ve{X}|)$ and $\partial_L |\ve{X}|$, correspondingly,
evaluated at $\ve{X}=\ve{X}_{\rm s}$.

It is important to note that if $\eta\neq 0$ the equations of motion of
a test particle contain not only the BD moments ${\cal M}_L$ and ${\cal
S}_L$ as in case of General Relativity, but also one more family of
mass moments ${\cal N}_L$ defined by (\ref{NL}).

\section{Translational and rotational equations of motion of $N$ extended bodies}
\label{translational-eqm}

Translational equations of motion of the origin of the local reference
system is given by (\ref{aei}).  In order to relate the origin of the
local reference system to the post-Newtonian center of mass of the body
(i.e., to fix $Q_a$) we can employ the local equations of motion in
the local PPN reference system

\begin{equation}\label{local-eq-m}
{\cal T}^{\alpha\beta}_{\ \ \ ;\beta}=0.
\end{equation}

\subsection{Local equations of motion: ${\cal T}^{0\beta}_{\ \ ;\beta}=0$}

Let us consider the integral

\begin{equation}\label{pN:fock:mass}
\int_{V} (-{\cal G})\,{\cal T}^{0\beta}_{\ \ ;\beta}\,d^3X=0.
\end{equation}

\noindent
In order to evaluate (\ref{pN:fock:mass}) we have to calculate the
Christoffel symbols as well as the metric determinant for the
local PPN metric

\begin{eqnarray}\label{G000}
\Gamma^0_{00}&=&-{1\over c^3}\,W_{,t}+\OO5,
\\
\label{G00a}
\Gamma^0_{0a}&=&-{1\over c^2}\,W_{,a}+\OO4,
\\
\label{G0ab}
\Gamma^0_{ab}&=&{1\over c^3}\,\left(
(1+\gamma)\,\left(W^a_{,b}+W^b_{,a}\right)+\gamma\,\delta^{ab}\,W_{,T}
\right)+\OO4,
\\
\label{-G}
-{\cal G}&=&1+{1\over c^2}\,2\,(3\gamma-1)\,W+\OO4.
\end{eqnarray}

\noindent
Substituting (\ref{G000})--(\ref{-G}) into (\ref{pN:fock:mass}) one
gets

\begin{equation}\label{pN:reqm:mass}
{d\over dT}\,M_T=F +\OO4,
\end{equation}

\noindent
where $M_T$ is the Tolman or ADM mass

\begin{eqnarray}\label{pN:mass:T}
M_T(T)&=&\int_V\,\rho_{_{\rm Tol}}(T,\ve{X})\, d^3X,
\end{eqnarray}

\noindent
with

\begin{eqnarray}\label{rhoT}
\rho_{_{\rm Tol}}(T,\ve{X})=
\Biggl(1+{1\over c^2}\biggl(&&
\left({1\over 2}+3(\gamma-1)\right)\,W_E
\nonumber\\
&&
+(3\,\gamma-2)\,\left(Q_a\,X^a+W_{\rm T}\right)
\biggr)
\Biggr)\,{1\over c^2}\,{\cal T}^{00}+\OO4.
\end{eqnarray}

\noindent
In case of an ideal fluid

\begin{eqnarray}\label{rhoT_ideal}
\rho_{_{\rm Tol}}(T,\ve{X})&=&
\hat\rho^*\left(1+{1\over c^2}
\left({1\over 2}\,V^2+\Pi-{1\over 2}\,W_{\rm E}
\right)\right)+\OO4,
\end{eqnarray}

\noindent
$\hat\rho^*$ being defined by (\ref{rho*-GRS}). This justifies that
$\rho_{_{\rm Tol}}$ as given by (\ref{rhoT}) is indeed the usual Tolman
density. However, the definition (\ref{rhoT}) is independent of the
model for ${\cal T}^{\mu\nu}$.  The right-hand side of
(\ref{pN:reqm:mass}) reads

\begin{eqnarray}\label{pN:reqm:mass:force}
F&=&{1\over c^2}\,\int_V
\Sigma^a\,\left(Q_a+W_{{\rm T},a}\right)\,d^3X+\OO4.
\end{eqnarray}

\noindent
For an isolated body the mass $M_T$ is obviously conserved.  The BD
mass ${\cal M}$ defined by (\ref{ML-BD-PPN}) is related with the Tolman
mass $M_T$ by

\begin{eqnarray}\label{MBD-MT}
{\cal M}=M_T&+&{1\over c^2}\,\eta\,\Omega_{\rm E}+
{1\over c^2}\,{1\over 6}\,(\gamma-1)\,
{d^2\over dT^2}\,\int_V\Sigma\,\ve{X}^2 d^3X
\nonumber \\
&+&
{1\over c^2} \left(1-\gamma-2\beta\right)\, Q_a
\int_V \Sigma\, X^a d^3X
\nonumber \\
&+&
{1\over c^2}\, \int_V \Sigma \left( (1-2\beta)\, W_{\rm T}-
\gamma\,W_{{\rm T},a}\, X^a\right) d^3X+\OO4,
\end{eqnarray}

\noindent
where

\begin{eqnarray}\label{OmegaE}
\Omega_{\rm E}&=&-{1\over 2}\,\int_V \Sigma\,W_{\rm E}\,d^3X+\OO2
\end{eqnarray}

\noindent
is the internal gravitational energy of the body.
One might have expected the appearance of the second term (containing
$\Omega_E$) in (\ref{MBD-MT}) since
$M_T$ is an inertial mass and ${\cal M}$ is the active gravitational
mass appearing in the expansion of the gravitational potential
(\ref{BD-expansion-PPN}) of the body. The third term can be
related with a Newtonian virial theorem. It
vanishes for an isolated body
under the assumption of secular stationarity. The last two terms in
(\ref{MBD-MT}) are proportional to $Q_a$ and the external tidal
potential $W_{\rm T}$ respectively, and vanish for an isolated body. We
see that the local equations of motion do not constrain the BD mass
${\cal M}$ as was the case in General Relativity, but only the Tolman mass
$M_T$. Moreover, in the framework of the PPN formalism the BD mass
${\cal M}$ is not conserved even for an isolated body unless some
specific physical conditions are met (secular stationarity and
$\dot\Omega_{\rm E}=0$).

\subsection{Local equations of motion:  ${\cal T}^{a\beta}_{\ \ ;\beta}=0$}

Let us now consider the integral

\begin{equation}\label{pN:fock:trans}
\int_{V} (-{\cal G}) \hat T^{a\beta}_{\ \ ;\beta}\,d^3X=0.
\end{equation}

The Christoffel symbols necessary for the evaluation of
(\ref{pN:fock:trans}) read

\begin{eqnarray}\label{Ga00}
\Gamma^a_{00}&=&-{1\over c^2}\,W_{,a}+
{2\over c^4}\left((\beta+\gamma)\,W\,W_{,a}
-(1+\gamma)\,W^a_{,T}\right)+\OO5,
\\
\label{Ga0b}
\Gamma^a_{0b}&=&{1\over c^3}\,\left(
(1+\gamma)\,\left(W^b_{,a}-W^a_{,b}\right)+
\gamma\,\delta^{ab}\,W_{,T}\right)+\OO5,
\\
\label{Gabc}
\Gamma^a_{bc}&=&{1\over c^2}\,\gamma\,\left(
-\delta^{bc}\,W_{,a}+\delta^{ab}\,W_{,c}+\delta^{ac}\,W_{,b}
\right)+\OO4,
\end{eqnarray}

\noindent
Substituting (\ref{Ga00})--(\ref{Gabc}) and (\ref{-G}) into
(\ref{pN:fock:trans}) one gets

\begin{equation}\label{pN:reqm:trans}
{d^2\over dT^2} M_T^a=F^a+{1\over c^2}\,F^a_\Psi+\OO4,
\end{equation}

\noindent
where

\begin{eqnarray}\label{pN:dipole:T}
M^a_T&=&\int_V\,\rho_{_{\rm Tol}}(T,\ve{X})\,X^a\,d^3X,
\end{eqnarray}

\begin{eqnarray}\label{Fa}
F^a&=&\int_V f^a\,d^3X+{1\over c^2}\,{d\over dT}\,\int_V g^a\,d^3X+\OO4,
\end{eqnarray}

\begin{eqnarray}\label{PPN:fa}
f^a=\Sigma\,(Q^a+W_{{\rm T},a})
+{1\over c^2}\,2(1+\gamma)\,\biggl(&&\Sigma\,
\left({1\over 2}\,\varepsilon_{abc}\,\dot C_b\,X^c+W^a_{{\rm T},T}\right)
\nonumber \\
&&
+\Sigma^b\,
\left(-\varepsilon_{abc}\,C_c+
W^a_{{\rm T},b}-W^b_{{\rm T},a}\right)
\biggr),
\end{eqnarray}

\begin{eqnarray}\label{PPN:ga}
g^a&=&\Sigma^b\,(Q_b+W_{\rm T,b})\,X^a
-(2\gamma+1)\,\Sigma^a\,(Q_b\,X^b+W_{\rm T}),
\end{eqnarray}

\begin{eqnarray}\label{Fa:Psi}
F^a_\Psi=\int_V \Sigma\,\Psi_{,a}\,d^3X+\OO2
=\eta\,\Omega_{\rm E}\,R^a_{\ i}\,a_E^i+\OO2.
\end{eqnarray}

It is easy to see from (\ref{pN:reqm:trans})--(\ref{Fa:Psi}) that for
isolated body $\ddot M_T^a$ vanishes. The BD mass dipole ${\cal M}^a$
defined by (\ref{ML-BD-PPN})and $M^a_T$ are related by

\begin{eqnarray}\label{MTa-MBDa}
{\cal M}^a&=&M^a_T+{1\over c^2}\,\eta\,\Omega_E^a
\nonumber \\
&&+{1\over c^2}\,{1\over 5}\,(\gamma-1)\,{d^2\over dT^2}\,
\int_V \Sigma\,\ve{X}^2\,X^a\,d^3X
-{1\over c^2}\,{1\over 2}\,(\gamma-1)\,{d\over dT}
\int_V \Sigma^a\,\ve{X}^2\,d^3X
\nonumber \\
&&+{1\over c^2}\,(1-2\beta)
\int_V\Sigma\,\left(Q_a X^a+W_{\rm T}\right)\,X^a\,d^3X+
\nonumber \\
&&+{1\over c^2}\,{1\over 2}\,\gamma\,
\int_V\Sigma\,\left(Q_a+W_{{\rm T},a}\right)\,\ve{X}^2\,d^3X
\nonumber \\
&&-{1\over c^2}\,\gamma\,
\int_V\Sigma\,\left(Q_b+W_{{\rm T},b}\right)\,X^b\,X^a\,d^3X
+\OO4,
\end{eqnarray}

\noindent
where

\begin{eqnarray}\label{OmegaEa}
\Omega_{\rm E}^a=-{1\over 2}\,\int_V \Sigma\, W_{\rm E}\, X^a\, d^3X+\OO2.
\end{eqnarray}

\noindent
If $\eta\neq0$ or $\gamma\neq1$, the second time derivative of the BD
mass dipole $\ddot {\cal M}^a$ does not vanish even for one isolated
body unless specific physical conditions are met (secular stationarity
and $\ddot\Omega_{\rm E}^a$).

\subsection{Local equations of motion:
$\varepsilon_{abc}\,X^b\,{\cal T}^{c\beta}_{\ \ ;\beta}=0$}
\label{Angular-momentum}

The local PPN reference system of an extended massive body described
above allows us to derive also rotational equations of motion of the
body in the framework of the PPN formalism.  Substituting
(\ref{Ga00})--(\ref{Gabc}) and (\ref{-G}) into

\begin{equation}\label{pN:fock}
\varepsilon_{abc}\,\int_{V} (-{\cal G})\,X^b\,{\cal T}^{c\beta}_{\ \ ;\beta}
\,d^3X=0
\end{equation}

\noindent
one gets the following rotational equations of motion
(see, \cite{dsx:93,klio:96} for a more detailed discussion)

\begin{equation}\label{dSdt-ppn}
{d\over dT}\,S^a=L^a+{1\over c^2}\,L^a_{\Psi}
+{\cal O}(c^{-4}),
\end{equation}

\noindent
where the PPN spin $S^a$ is defined by

\begin{eqnarray}\label{pN:spin}
S^a&=&\varepsilon_{abc} \int_{V} X^b\, p^c\, d^3X
+\OO4
\end{eqnarray}

\noindent
with

\begin{eqnarray}\label{pN:Q}
p^a&=&\Sigma^a\, (1+ {5\gamma-1\over c^2}\, W)
-{1\over 2c^2}\,G\,\Sigma
\int_{V} \Sigma^b(T,\ve{X}')\,\,{(4\gamma+3)\,\delta^{ab} +
n^a n^b \over |\ve{X}-\ve{X}'|}\,\, d^3X'+{\cal O}(c^{-4}),
\end{eqnarray}

\begin{eqnarray}\label{ni}
n^a&=&{X^a-X'^a\over |\ve{X}-\ve{X}'|}.
\end{eqnarray}

\noindent
Here,

\begin{eqnarray}\label{La}
L^a&=&\varepsilon_{abc}\,\int_{V} X^b\,f^c\,d^3X+\OO4,
\end{eqnarray}

\noindent
and

\begin{eqnarray}\label{torque:Psi}
L_\Psi^a&=&\varepsilon_{abc}\,\int_V \Sigma\,X^b\,\Psi_{,c}\, d^3X=
\eta\,\varepsilon_{abc}\,\Omega_{\rm E}^b\,R^c_{\ i}\,a_E^i+\OO4,
\end{eqnarray}

\noindent
where $f^a$ and $\Omega_{\rm E}^a$ are defined by (\ref{PPN:fa}) and
(\ref{OmegaEa}), respectively. For one isolated body
$S^a$ is conserved. It is easy to see that $S^a$ coincide with the
BD spin moment ${\cal S}^a$ defined by (\ref{SL-BD-PPN}) in the Newtonian
approximation

\begin{eqnarray}\label{Sa-cal-Sa}
S^a={\cal S}^a+\OO2.
\end{eqnarray}

The torque $L^a_\Psi$ comes from the term $\Psi$ in the local PPN
metric and represents an analogy of the Nordtvedt effect in the
rotational equations of motion of an extended body in the local PPN
metric. These rotational equations of motion and the torque
$L_\Psi^a$ has been discussed in a previous paper
\cite{klioner:soffel:1998}.

\subsection{Multipole expansions of the relations between the Tolman and
BD masses and mass dipoles}

For further analysis it is interesting to expand explicitly the relations
for the Tolman mass $M_T$ and mass dipole $M_T^a$ with the
BD mass ${\cal M}$ and mass dipole ${\cal M}^a$ defined by
(\ref{MBD-MT}) and (\ref{MTa-MBDa}) respectively. One gets

\begin{eqnarray}\label{MBD-MT-expanded}
{\cal M}=M_T&+&{1\over c^2}\,\eta\,\Omega_E+
{1\over c^2}\,{1\over 6}\,(\gamma-1)\,
\ddot{\cal N}
\nonumber \\
&+&
{1\over c^2} \sum_{l=1}^\infty\,{1\over l!}\,
\left(1-2\beta-\gamma\,l\right) {\cal M}_L\,Q_L+\OO4,
\end{eqnarray}

\begin{eqnarray}\label{MBDa-MTa-expanded}
{\cal M}^a&=&M^a_T+{1\over c^2}\,\eta\,\Omega_{\rm E}^a
\nonumber \\
&&
-{1\over c^2}\,{1\over 10}\,(\gamma-1)\,
\,\ddot{\cal N}_a
+{1\over c^2}\,{3\over 5}\,(\gamma-1)\,\dot{\cal P}_a
\nonumber \\
&&+{1\over c^2}\,
\sum_{l=1}^\infty\,{1\over l!}\,
\left(1-2\beta-\gamma\,l\right)\,
{\cal M}_{aL}\,Q_L+
\nonumber \\
&&+{1\over c^2}\,
\sum_{l=0}^\infty\,{1\over l!}\,
\left({1\over 2}\,\gamma-2\,\beta+1\right)\,{1\over 2l+3}
\,{\cal N}_{L}\,Q_{aL}
+\OO4,
\end{eqnarray}

\noindent
where

\begin{eqnarray}\label{P}\label{Pa}
{\cal P}_a&=&\int_V \Sigma^b\,\hat X^{ab}\,d^3X.
\end{eqnarray}

\noindent
Here, ${\cal N}_L$ is defined by (\ref{NL}). The mass and mass dipole
are the only moments which appear in two different forms in the
formalism: 1) the BD mass and BD mass dipole in the multipole
expansions of the internal gravitational potential
(\ref{BD-expansion-PPN}), and 2) the Tolman mass and Tolman mass
dipole in the left-hand side of the local equations of motions
(\ref{pN:reqm:mass}) and (\ref{pN:reqm:trans}).  The physical significance
of the BD moments lies in their appearance in the multipole expansion of
the post-Newtonian gravitational field, that of the
Tolman mass and mass dipole lies in representing integrals of
local equations of motion for one isolated body. In General Relativity
the BD and Tolman mass and mass dipoles coincide for isolated bodies.
In contrast to this, within the PPN formalism this is no longer the
case and the BD mass ${\cal M}$ and mass dipole ${\cal M}^a$ and the
Tolman mass $M_T$ and mass dipole $M^a_T$ are essentially different
objects.  Higher multipole moments appear only in the multipole
expansions of the internal gravitational potential
(\ref{BD-expansion-PPN}).

\subsection{Multipole-expanded local translational equations of motion}
\label{LEQM-expanded}

Using the definitions of the BD multipole moments and external tidal moments
introduced in Sections \ref{Section-multipoles-internal} and
\ref{Section-multipoles-external} one can derive the following
multipole expansions

\begin{eqnarray}\label{dotM-BD-mutipoles}
\dot{\cal M}&=&\Delta \dot M
\nonumber \\
&&-{1\over c^2} \sum_{l=1}^\infty {1\over l!}\,
\biggl(
\Bigl(l+1+(\gamma-1)\,l+2\,(\beta-1)\Bigr)\, {\cal M}_L \dot Q_L
\nonumber \\
&&
\phantom{{1\over c^2} \sum_{l=1}^\infty {1\over l!}\,\biggl(}
+
\Bigl(l+(\gamma-1)\,l\,+2\,(\beta-1)\Bigr)\,\dot{\cal M}_L Q_L
\biggr)+\OO4,
\end{eqnarray}

\begin{eqnarray}\label{dotMa-BD-mutipoles}
\ddot{\cal M}_a&=&\Delta \ddot M_a + R^a_{\ i}\,a_E^i\,\Delta M
\nonumber \\
&&+\sum_{l=0}^\infty {1\over l!}\,{\cal M}_{L}\,Q_{aL}
\nonumber \\
&&
-{1\over c^2}\,2\,(1+\gamma)\,\sum_{l=0}^\infty{1\over l!}\,
\biggl(
{1\over l+2}\,{\cal S}_{L+1}\,C_{aL+1}
+{1\over l+2}\,\varepsilon_{abc}\,{\cal M}_{bL}\,\dot C_{cL}
+{1\over l+1}\,\varepsilon_{abc}\,\dot{\cal M}_{bL}\,C_{cL}
\nonumber\\
&&
\phantom{-{1\over c^2}\,2\,(1+\gamma)\,\sum_{l=0}^\infty{1\over l!}\,\biggl(}
+{(l+1)\over(l+2)^2}\,\varepsilon_{abc}\,{\cal S}_{bL}\,\dot Q_{cL}
+{1\over l+2}\,\varepsilon_{abc}\,\dot {\cal S}_{bL}\,Q_{cL}
\biggr)
\nonumber \\
&&
-{1\over c^2}\sum_{l=1}^\infty{1\over l!}\,
\left(
% factoring
%{2l^3+7l^2+15l+6\over (l+1)\,(2l+3)}
{(2l+1)\,(l^2+3l+6)\over (l+1)\,(2l+3)}
+(\gamma-1)\,{2l^3+5l^2+7l+2\over (l+1)\,(2l+3)}
+2\,(\beta-1)
\right)
\,{\cal M}_{aL}\,\ddot Q_L
\nonumber \\
&&
-{1\over c^2}\sum_{l=1}^\infty{1\over l!}\,
\left(
% factoring
%{2l^3+5l^2+12l+5\over (l+1)^2}
{(2l+1)\,(l^2+2l+5)\over (l+1)^2}
+2\,(\gamma-1)\,{l^3+2l^2+3l+1\over (l+1)^2}
+4\,(\beta-1)
\right)
\,\dot{\cal M}_{aL}\,\dot Q_L
\nonumber \\
&&
-{1\over c^2}\sum_{l=1}^\infty{1\over l!}\,
\left(
{l^2+l+4\over l+1}+
(\gamma-1)\,{l^2+l+2\over l+1}+2\,(\beta-1)
\right)
\,\ddot{\cal M}_{aL}\,Q_L
\nonumber \\
&&
-{1\over c^2}\,\eta\,\sum_{l=0}^\infty{1\over l!}\,
{1\over 2l+3}\,{\cal M}_{aL}\,R^b_{\ i}\,a_E^i\,Q_{bL}
\nonumber \\
&&
-{1\over 2c^2}\,\eta\,\sum_{l=0}^\infty{1\over l!}\,
{1\over 2l+3}\,R^b_{\ i}\,a_E^i\,{\cal N}_{L}\,Q_{abL}
\nonumber \\
&&
-{1\over 2c^2}\,\eta\,\sum_{l=0}^\infty{1\over l!}\,
{1\over 2l+3}\,
{d^2\over dT^2}\,\biggl( {\cal N}_L\,Q_{aL} \biggr)
\nonumber \\
&&
-{1\over 6c^2}\,(\gamma-1)\,{d^2\over dT^2}\,
\left({\cal N}\,R^a_{\ i}\,a_E^i\right)
\nonumber \\
&&
+{1\over 3c^2}\,{\cal M}_a\,\left(
(\gamma-1)\,\ddot{\overline U}(t,\ve{x}_E)
-2\,(\beta-1)\,a_E^2
\right)
+{2\over 5c^2}\,(\gamma-1)\,R^b_{\ i}\,\ddot a_E^j\,{\cal M}^{ab}
\nonumber \\
&&
+{1\over c^2}\,(\gamma-1)\,
\left(
{1\over 2}\,\dot{\cal M}_{ab}\,R^b_{\ i}\,\dot a_E^i
-{1\over 2}\,\varepsilon_{abc}\,{\cal S}_b\,R^c_{\ i}\,\dot a_E^i
\right)
+\OO4,
\end{eqnarray}

\begin{eqnarray}\label{dotS-Newton}
\dot{\cal S}_a&=&\sum_{l=0}^{\infty} {1\over l!} \,
\varepsilon_{abc}\,{\cal M}_{bL}\,Q_{cL}+\OO2.
\end{eqnarray}

\noindent
with

\begin{equation}\label{Delta-M}
\Delta M={1\over c^2}
\left(\eta\,\Omega_E+{1\over 6}\,(\gamma-1)\,\ddot {\cal N}\right)
\end{equation}

\noindent
and

\begin{equation}\label{Delta-Ma}
\Delta M_a=
{1\over c^2}\left(\eta\,\Omega_E^a
-{1\over 10}\,(\gamma-1)\,\ddot {\cal N}_a
+{3\over 5}\,(\gamma-1)\,\dot{\cal P}_a\right).
\end{equation}

\noindent
$\Delta M$ and $\Delta M_a$ represent the internal parts of the
difference between the gravitational BD mass and mass dipole,  ${\cal
M}$ and ${\cal M}^a$, and the Tolman mass $M_T$ and $M_T^a$ given by
(\ref{MBD-MT-expanded})--(\ref{MBDa-MTa-expanded}). In the limit of
General Relativity (\ref{dotM-BD-mutipoles})--(\ref{dotS-Newton})
coincide with Eqs. (4.20a)--(4.21c) of \cite{dsx:92}.

Analogous to the translational equations of motion discussed in Section
\ref{section:eqm-test-particle}, Eqs.
(\ref{dotM-BD-mutipoles})-(\ref{dotMa-BD-mutipoles}) contain not only
the two families of BD moments ${\cal M}_L$ and ${\cal S}_L$, but one
more family of multipole moments ${\cal N}_L$.  Moreover,
(\ref{dotM-BD-mutipoles})-(\ref{dotMa-BD-mutipoles}) contain also a
number of additional terms that vanish in General Relativity, but do
not vanish for $\eta=0$. The physical meaning of some of these terms
will be discussed in Section \ref{Monopole-spin-dipole-approximation}
below.

\subsection{Multipole-expanded rotational equations of motion}
\label{rotational-eqm}

In the same way for the post-Newtonian rotational equations of motion one
gets

\begin{eqnarray}\label{dotSa-BD-mutipoles}
\dot S^a&=&\varepsilon_{abc}\,\sum_{l=0}^\infty{1\over l!}
\left({\cal M}_{bL}\,Q_{cL}-
{1\over c^2}\,2(1+\gamma)\,{l+1\over l+2}\,{\cal S}_{bL}\,C_{cL}\right)
\nonumber \\
&&
+{1\over c^2}\,L^a_\Delta
+{1\over c^2}\,{d\over dT}\,S_1^a
+{1\over c^2}\,{d\over dT}\,S_2^a
+{1\over c^2}\,L_\eta^a
+\OO4
\end{eqnarray}

\noindent
with

\begin{eqnarray}\label{LDelta}
L_\Delta^a&=&L^a_\Psi
-{1\over 10}\,(\gamma-1)\,
\varepsilon_{abc}\,\ddot {\cal N}_b\,R^c_{\ i}\,a_E^i
+{3\over 5}\,(\gamma-1)\,
\varepsilon_{abc}\,\dot {\cal P}_b\,R^c_{\ i}\,a_E^i=
c^2\,\varepsilon_{abc}\,\Delta M_b\,R^c_{\ i}\,a_E^i,
\end{eqnarray}

\begin{eqnarray}\label{L1}
S_1^a&=&-2(1+\gamma)\,\sum_{l=1}^\infty {1\over l!}\,{l\over l+1}\,
{\cal M}_{aL}\,C_L,
\end{eqnarray}

\begin{eqnarray}\label{L2}
S_2^a&=&2(1+\gamma)\,\sum_{l=0}^\infty {1\over l!}\,{1\over 2l+3}
{\cal N}_L\,C_{aL}
\nonumber \\
&&+2(1+\gamma)\,\varepsilon_{abc}\,\sum_{l=0}^\infty{1\over l!}\,
{(2l+3)\over(l+2)(2l+5)}\,{\cal P}_{bL}\,Q_{cL}
\nonumber \\
&&+\varepsilon_{abc}\,\sum_{l=0}^\infty{1\over l!}\,
{l+2(\gamma+2)\over 2(l+2)(2l+5)}\,
\left({\cal N}_{bL}\,\dot Q_{cL}-\dot {\cal N}_{bL}\,Q_{cL}\right)
\nonumber \\
&&+(1+\gamma)\,\varepsilon_{abc}\,\sum_{l=0}^\infty{1\over l!}\,
{1\over(l+2)(2l+5)}\,{d\over dT}\left({\cal N}_{bL}\,Q_{cL}\right)
\nonumber \\
&&
-{3\over 5}\,(\gamma-1)\,\varepsilon_{abc}\,{\cal P}_b\,R^c_{\ i}\,a_E^i
\nonumber \\
&&
-{1\over 5}\,(\gamma-1)\,\varepsilon_{abc}\,
\left({\cal N}_b\,R^c_{\ i}\,\dot a_E^i
-\dot{\cal N}_b\,R^c_{\ i}\,a_E^i\right)
\nonumber \\
&&
-{1\over 10}\,(\gamma-1)\,\varepsilon_{abc}\,
{d\over dT}\,\left({\cal N}_b\,R^c_{\ i}\,a_E^i\right)
\end{eqnarray}

\noindent
and

\begin{eqnarray}\label{L3}
L_\eta^a&=&-\eta\,\varepsilon_{abc}\,\sum_{l=0}^\infty{1\over l!}
\,{1\over2(2l+5)}\,{\cal N}_{bL}\,R^d_{\ i}\,a_E^i\,Q_{cdL}.
\end{eqnarray}

\noindent
Here, $L^a_\Psi$ is defined by (\ref{torque:Psi}) and
${\cal P}_L$ by (\ref{PL}). We prefer here to
consider $L^a_\Psi$ as part of $L^a_\Delta$ which is proportional to
the difference of the BD and Tolman dipoles $\Delta M_a$.  This might
provide a deeper insight into the nature of this torque.  As in
General Relativity we can now change the definition of the
post-Newtonian spin into $S'^a=S^a+S_1^a+S_2^a$ to get rid of $S_1^a$
and $S_2^a$ on the right-hand side of (\ref{dotSa-BD-mutipoles}) (see,
\cite{dsx:93} for a discussion of this point):

\begin{eqnarray}\label{S'}
\dot S'^a=
\varepsilon_{abc}\,\sum_{l=0}^\infty{1\over l!}
\left({\cal M}_{bL}\,Q_{cL}-
{1\over c^2}\,2(1+\gamma)\,{l+1\over l+2}\,{\cal S}_{bL}\,C_{cL}\right)
+{1\over c^2}\,\left(L^a_\Delta+L_\eta^a\right)
+\OO4.
\end{eqnarray}

\noindent
The torque $L^a_\eta$ is in some sense similar to $L^a_\Psi$: it is
proportional to the Nordtvedt parameter $\eta$ and the acceleration of
the body relative to the global reference system $a_E^i$. For that
reason, it is related with a violation of the Strong Equivalence
Principle. $L^a_\Psi$ has been discussed in detail in a previous
paper \cite{klioner:soffel:1998}.

Let us note here that the rotational equations of motion and the
definition of the post-Newtonian spin should be considered as being
formal in the first place.  Further analyses will be needed to
see how these concepts can be used efficiently in practice.

\subsection{Mass monopole-spin dipole approximation}
\label{Monopole-spin-dipole-approximation}

Let us discuss the skeletonized equations of motion of $N$ extended
massive bodies with full multipole structure derived above for the case
that the $N$ bodies possess only mass and spin.  The corresponding mass
monopole-spin dipole model for the multipole structure of each body $B$ can
be mathematically formulated as

\begin{eqnarray}\
\label{md:ML=0}
{\cal M}_L&=&\OO4,\qquad l\ge 1,
\\ \label{md:SL=0}
{\cal S}_L&=&\OO2,\qquad l\ge 2,
\\ \label{md:NL=0}
{\cal N}_L&=&\OO2,\qquad l\ge 1,
\\ \label{md:Pa=0}
{\cal P}_a&=&\OO2,
\\ \label{md:Omegaa=0}
\Omega^a&=&\OO2.
\end{eqnarray}

\noindent
These assumptions agree with what we expect from a naive Newtonian
model of spherically symmetric rotating body.  From
(\ref{MBD-MT-expanded})--(\ref{MBDa-MTa-expanded}),
(\ref{dotM-BD-mutipoles})--(\ref{dotMa-BD-mutipoles}),
(\ref{dotSa-BD-mutipoles})--(\ref{L3}) and
(\ref{md:ML=0})--(\ref{md:Omegaa=0}) one finds

\begin{equation}\label{md:MT-const-point}
\dot M_T=\OO4,
\end{equation}

\begin{equation}\label{md:MT-MBD-point}
{\cal M}=M_T+\Delta M+\OO4,
\end{equation}

\begin{equation}\label{md:MTi-MBDi-point}
{\cal M}^a=M_T^a+\OO4,
\end{equation}

\begin{eqnarray}\label{Qa-md}
&&\Delta M\,R^a_{\ i}\,a_E^i+
{\cal M}\, Q_a
-{1\over 6c^2}\,\eta\,{\cal N}\,R^b_{\ i}\,a_E^i\,Q_{ab}
+{1\over 6c^2}\,(1-\gamma)\,{d^2\over dT^2}
\left({\cal N}\,R^a_{\ i}\,a_E^i\right)
\nonumber \\
&&
-{1\over c^2}\,(1+\gamma)\,{\cal S}_b\,C_{ab}
-{1\over c^2}\,{1\over 2}\,(\gamma-1)\,
\varepsilon_{abc}\,{\cal S}_b\,R^c_{\ i}\,\dot a_E^i
=\OO4.
\end{eqnarray}

\noindent
Eq. (\ref{Qa-md}) allows us to derive $Q_a$ for the case that the origin
of the local reference system coincides with the BD center of mass of
the central body (${\cal M}_a\equiv0$). For convenience we split $Q_a$
into a part independent of the spin of the body $Q_a^{\rm m}$ (that
is, $Q_a^{\rm m}$ corresponds to mass monopole model for the central body)
and a part proportional to the spin of the body $Q_a^{\rm s}$

\begin{eqnarray}\label{md-Qa}
Q_a&=&Q_a^{\rm m}+Q_a^{\rm s},
\\ \label{mono-Qa}
\nonumber\\
Q_a^{\rm m}&=&
-{\Delta M\over {\cal M}}\, R^a_{\ i}\,a_E^i
+{1\over 6c^2}\,\eta\,{{\cal N}\over {\cal M}}\,R^b_{\ i}\,a_E^i\,Q_{ab}
\nonumber\\
&&-{1\over 6c^2}\,(1-\gamma)\,{1\over {\cal M}}\,{d^2\over dT^2}
\left({\cal N}\,R^a_{\ i}\,a_E^i\right)+
\OO4,
\\ \label{spin-Qi}
\nonumber\\
Q_a^{\rm s}&=&
{1\over c^2}\,(1+\gamma)\,{1\over {\cal M}}\,{\cal S}_b\,C_{ab}
-{1\over 2c^2}\,(1-\gamma)\,{1\over {\cal M}}\,
\varepsilon_{abc}\,{\cal S}_b\,R^c_{\ i}\,\dot a_E^i+\OO4.
\end{eqnarray}

\noindent
From assumptions (\ref{md:ML=0})--(\ref{md:SL=0}) and expansions
(\ref{BD-expansion-PPN})--(\ref{BD-expansion-PPN-Wi})
in the local PPN reference system of each body $B$ we have

\begin{eqnarray}\label{md-local-WB}
W_B&=&G\;{\cal M}_B\;{1\over |\ve{X}_B|}+\OO4,
\\
\label{md-local-WBa}
W^a_B&=&{1\over 2}\,G\,\varepsilon_{abc}\,{\cal S}_B^c\,
\partial_b\,{1\over |\ve{X}_B|}+\OO2.
\end{eqnarray}

\noindent
Here, for each body the skeletonized-harmonic gauge (\ref{kill-Lambda-PPN})
was used (see, \cite{dsx:91} for more detail). $\ve{X}_B$ are spatial
coordinates of the local PPN reference system of body $B$.
Now we derive the potentials of body $B$ in the global PPN metric
following the steps of Appendix C of \cite{dsx:91}. From
(\ref{trans:space}), (\ref{D^{ij}}) and (\ref{D^{ijk}}) one gets

\begin{equation}\label{md:1/X-1/rb}
{1\over |\ve{X}_B|}={1\over r_B}\,\left(
1-{1\over c^2}\left({1\over 2}\left({\dot x_B^in_B^i}\right)^2+
\gamma\,\overline w_B(\ve{x}_B)+
{1\over 2}\gamma\,\ddot x_B^ir_B^i
\right)
\right)+\OO4,
\end{equation}

\noindent
where $r_B^i=x^i-x_B^i$, $n_B^i=r_B^i/r_B$. Here, $x_B^i$ are the
coordinates of body $B$ in the global PPN reference system, and
$\overline w_B$ is the external potential appearing, e.g., in
(\ref{BRS:split}) when constructing the local PPN reference system for
body $B$. Then substituting (\ref{md-local-WB})--(\ref{md-local-WBa})
and (\ref{md:1/X-1/rb}) into (\ref{g00:c2+c4:matching}) and
(\ref{W_E^a}) we get the explicit form of the gravitational potential
of the body $B$ in the global PPN reference system

\begin{eqnarray}\label{md-wB}
w_B={G{\cal M}_B\over r_B}\,\Biggl(
1+{1\over c^2}\Biggl(&&
\left(\gamma+{1\over 2}\right) \dot x_B^2-(2\beta-1)\,\overline w_B(\ve{x}_B)
\nonumber \\
&&+{1\over 2}\left(\dot x_B^2-\ddot x_B^i r_B^i
-\left(\dot x_B^i n_B^i\right)^2\right)
\Biggr)
\Biggr)
\nonumber \\
&&
\hskip -3cm
+{1\over c^2}\,(1+\gamma)\,G\,\varepsilon_{ijk}\,\dot x_B^i\,
s_B^k\,\partial_j\,{1\over r_B}
+\OO4.
\end{eqnarray}

\noindent
Here, $s_B^i=R^{a\,(B)}_{\,i}\,S_B^a$ and $R^{a\,(B)}_{\,i}$ is the
rotational matrix $R^a_{\ i}$ appearing in the local PPN reference
system for body $B$. The expression for the vector potential
$W_B^a$ in the
global reference system can be derived immediately from
(\ref{W_E^a}), (\ref{md-wB}) and (\ref{md-local-WBa})

\begin{equation}\label{md-global-wBi}
w_B^i=
{G\,{\cal M}_B\over r_B}\,\dot x_B^i
+{1\over 2}\,G\,\varepsilon_{ijk}\,
s_B^k\,\partial_j\,{1\over r_B}
+\OO2.
\end{equation}

\noindent
Now we note that in analogy to (\ref{BRS:split}) the gravitational
potentials of the global PPN metric can be split into the sum of the
contributions of the bodies

\begin{eqnarray}\label{md-w+wB-split}
&&w=\sum_B w_B +\OO4,
\nonumber\\
&&w^i=\sum_B w^i_B+\OO2.
\end{eqnarray}

\noindent
Hence we get the explicit form of the gravitational potentials of the
global PPN reference system within our mass monopole-spin dipole model
(\ref{md-local-WB})--(\ref{md-local-WBa})

\begin{eqnarray}\label{md-w}
w&=&\sum_B {G{\cal M}_B\over r_B}\,\left(
1+{1\over c^2}\left(
\left(\gamma+{1\over 2}\right) \dot x_B^2-(2\beta-1)
\sum_{C\ne B} {G{\cal M}_C\over r_{CB}}\right)\right)
\nonumber\\
&&+{1\over c^2}\,{1\over 2}\sum_B G{\cal M}_B\, r_{B,tt}
\nonumber\\
&&
+\,{1\over c^2}\,(1+\gamma)\,\sum_B G
\,\varepsilon_{ijk}\,\dot x_B^i\,s_B^k\,\partial_j\,{1\over r_B}
+\OO4,
\\ \label{md-wi}
w^i&=&\sum_B {G{\cal M}_B\over r_B} \dot x_B^i
+{1\over 2}\,\sum_B G\,\varepsilon_{ijk}\,s_B^k\,\partial_j\,{1\over r_B}
+\OO2,
\end{eqnarray}

\noindent
where $\ve{r}_{CB}=\ve{x}_C-\ve{x}_B$ and $r_{CB}=|\ve{r}_{CB}|$ for
any $C$ and $B$.

The equations of translational motion of body $E$ (\ref{aei}) can be now
written as

\begin{eqnarray}\label{md-a_E}
a_E^i=a_{E,\rm m}^i+a_{E,\rm s}^i.
\end{eqnarray}

\noindent
Here the part $a_{E,\rm m}^i$ depends only upon the masses of the bodies

\begin{eqnarray}\label{mono-aEi}
a_{E,\rm m}^i&=&-\sum_{B\ne E} G\,{\cal M}_B\,
{r_{EB}^i\over r_{EB}^3}
-R^a_{\ i}\,Q^a_{\rm m}
\nonumber\\
&&+{1\over c^2}\,\sum_{B\ne E} G\,{\cal M}_B\,
{r_{EB}^i\over r_{EB}^3}\,
\Biggl\{
(2\beta-1) \sum_{C\ne B} {G\,{\cal M}_C\over r_{BC}}
+2(\gamma+\beta) \sum_{C\ne E} {G\,{\cal M}_C\over r_{EC}}
+2(1+\gamma) {G\,{\cal M}_E\over r_{EB}}
\nonumber \\
&&
\phantom{
+{1\over c^2}\,\sum_{B\ne E} G\,{\cal M}_B\,
{r_{EB}^i\over r_{EB}^3}\,
\Biggl\{ }
+{3\over 2} {{\left(r_{EB}^j \dot x_B^j\right)}^2\over r_{EB}^2}
-{1\over 2}\sum_{C\ne E,B} G\,{\cal M}_C\,{r_{EB}^j\,r_{BC}^j\over r_{BC}^3}
\nonumber \\
&&
\phantom{
+{1\over c^2}\,\sum_{B\ne E} G\,{\cal M}_B\,
{r_{EB}^i\over r_{EB}^3}\,
\Biggl\{ }
-(1+\gamma)\,\dot x_B^j\,\dot x_B^j
-\gamma\,\dot x_E^j\,\dot x_E^j
+2(1+\gamma)\,\dot x_E^j\,\dot x_B^j
\Biggr\}
\nonumber \\
&&+{1\over c^2}\,\sum_{B\ne E} G\,{\cal M}_B\,
{r_{EB}^j\over r_{EB}^3}\,
\biggl\{
 2(1+\gamma)\,\dot x_E^j-(2\gamma+1)\,\dot x_B^j\biggr\}\,
(\dot x_E^i-\dot x_B^i)
\nonumber \\
&&-{1\over c^2}\,\left(2\gamma+{3\over 2}\right)\,
\sum_{B\ne E}{G\,{\cal M}_B\over r_{EB}}\,
\sum_{C\ne E,B} G\,{\cal M}_C\,{r_{BC}^i \over r_{BC}^3}+\OO4.
\end{eqnarray}

\noindent
These are the PPN equations of motion for a system of
mass monopoles. They agree with the well-known PPN
Lorentz-Droste-Einstein-Infeld-Hoffmann equations of motion (see,
e.g., \cite{will:93}).  Finally, $a_{E,\rm s}^i$ are additional
acceleration terms due to the spins of the bodies

\begin{eqnarray}\label{spin-aEi}
a_{E,\rm s}^i&=&-R^a_{\ i}\,Q^a_{\rm s}
\nonumber\\
&&
+{1\over c^2}\,(1+\gamma)\,G\,\sum_{B\neq E}\,
\left(\varepsilon_{ijk}\,s_B^k\,(\dot x_E^l-\dot x_B^l)\,
\partial_{jl}\,{1\over r_{EB}} +
\varepsilon_{jlk}\,s_B^k\,(\dot x_E^l-\dot x_B^l)\,
\partial_{ji}\,{1\over r_{EB}}
\right)
\nonumber\\
&&
+\OO4.
\end{eqnarray}

\noindent
The geodetic deviation
$Q^a_{\rm s}$ due to spin of body E can be further simplified with the
mass monopole-spin dipole model we consider. From (\ref{C_L}),
(\ref{md-w}) and (\ref{md-wi}) one gets

\begin{eqnarray}\label{spin-Qa-md}
R^a_{\ i}\,Q^a_{\rm s}&=&
-{1\over c^2}\,{1\over {\cal M}_E}\,
\varepsilon_{ijk}\,s_E^j\,\dot a_E^k
\nonumber\\
&&-{1\over c^2}\,(1+\gamma)\,{1\over {\cal M}_E}
\,G\,\sum_{B\neq E}\,{\cal M}_B\,
\biggl(
\varepsilon_{jlk}\,s_E^k\,(\dot x_E^l-\dot x_B^l)\,
\partial_{ij}\,{1\over r_{EB}}
\nonumber\\
&&
\phantom{+{1\over c^2}\,(1+\gamma)\,{1\over {\cal M}_E}
\,G\,\sum_{B\neq E}\,{\cal M}_B\,\biggl(}
+\varepsilon_{ijk}\,s_E^k\,(\dot x_E^l-\dot x_B^l)\,
\partial_{jl}\,{1\over r_{EB}}
\biggr)
\nonumber\\
&&+{1\over 2c^2}\,(1+\gamma)\,{1\over {\cal M}_E}
\,G\,\sum_{B\neq E}\,s_E^j\,s_B^k\,\partial_{ijk}\,{1\over r_{EB}}
+\OO4.
\end{eqnarray}

\noindent
Here we used the identity $\varepsilon_{ij[k}\,A_{l]m}= {1\over
2}\,\varepsilon_{klm}\,A_{ij}$ valid of any trace-free $A_{ij}$
(so that, $A_{ii}=0$). Note that

\begin{eqnarray}\label{partial-ab}
\partial_{ij}\,{1\over r_{EB}}&\equiv&
{3\,r^i_{EB}\,r^j_{EB}-\delta^{ij}\,r^2_{EB} \over r^5_{EB}},
\\
\nonumber\\
\label{partial-abc}
\partial_{ijk}\,{1\over r_{EB}}&\equiv&
{3\,r^2_{EB}\,
(r^i_{EB}\,\delta^{jk}+r^j_{EB}\,\delta^{ik}+r^k_{EB}\,\delta^{ij})
-15\,r^i_{EB}\,r^j_{EB}\,r^k_{EB}
 \over r^7_{EB}}.
\end{eqnarray}

\noindent
Equations (\ref{md-a_E})--(\ref{partial-abc}) together with
(\ref{mono-Qa}) for $\gamma=\beta=1$ agree with Eqs. (6.29)--(6.36) of
\cite{dsx:92}. These equations of motion can be derived from Lagrangian
quoted in Section \ref{Introduction}.  Note that we did not assume the
constancy of the Blanchet-Damour masses ${\cal M}_B$.  Strictly
speaking one has to account for the time dependence of the masses
${\cal M}_B$ during a [numerical] integration of the equations of
motion (\ref{mono-aEi}). It is also clear that this effect is very
small and usually may be neglected in practical applications. Another
point is that the Nordtvedt effect here is represented here not only by
the terms $\eta\,\Omega_B$ and $\eta\,\Omega_E$, but through the whole
difference of the BD active gravitational mass ${\cal M}$ and the
inertial Tolman mass $M_T$ of each body.  According to
(\ref{md:MT-MBD-point}) and (\ref{Delta-M}) the difference contains not
only $\eta\,\Omega$, but also terms $(\gamma-1)\,\ddot{\cal N}/6$ that
do not vanish even for spherical bodies.

Taking into account that the BD mass ${\cal M}_B$ for each body is the
active gravitational mass (see, (\ref{MBD-MT})), it is easy to show
that these equations of motion coincide with Eqs. (6.31)--(6.34) and
(6.47) of \cite{will:93} if only the two PPN parameters $\beta$ and
$\gamma$ are retained there, all total derivatives of any internal
integrals are dropped in our equations (secular stationarity is assumed
in \cite{will:93}) and the ${\cal N}$-dependent terms are neglected in
our equations. The ${\cal N}$-dependent terms proportional to $\eta$
and to $\gamma-1$ can be explained by a coupling of the body's
extension to the background Ricci tensor, as already mentioned in
Section \ref{Introduction} and coincides with the corresponding terms
derived by Nordtvedt in
\cite{Nordtvedt:1971,Nordtvedt:1983,Nordtvedt:1991,Nordtvedt:1994}.

The spin-dependent part of the acceleration
(\ref{spin-aEi})--(\ref{spin-Qa-md}) coincides with the known
spin-orbit coupling derived, e.g., in
\cite{Damour:1982,Damour:Taylor:1992,bar:75} and can be derived from
Lagrangian (\ref{Lagr:spin}). The first term in (\ref{spin-Qa-md}) is
independent of the PPN parameters. It results from the well-known
Thomas precession and has been first discussed by Damour
\cite{Damour:1982} (see, also \cite{Damour:Taylor:1992}).

For the mass monopole-spin dipole model (\ref{md:ML=0})--(\ref{md:Omegaa=0})
extended by assuming

\begin{equation}\label{PL=0}
{\cal P}_L=\OO2,\quad l\ge2
\end{equation}

\noindent
the rotational equations of motion (\ref{dotSa-BD-mutipoles}) read

\begin{eqnarray}\label{dotSa-monopole-dipole}
\dot S^a&=&-{1\over c^2}\,(1+\gamma)\,\varepsilon_{abc}\, {\cal
S}_b\,C_c
+{1\over c^2}\,{2\over 3}\,(1+\gamma)\,{d\over dT}\,\left(N\,C_a\right)
+\OO4.
\end{eqnarray}

\noindent
The second term in the right-hand side of (\ref{dotSa-monopole-dipole})
can be absorbed by changing the definition of the post-Newtonian spin
as it was done in \cite{dsx:93} for General Relativity.

If the local reference system is chosen to be kinematically
nonrotating, $R^a_{\,i}=\delta^a_{\,i}$, then the first term in
(\ref{dotSa-monopole-dipole}) gives precession of the spin due to
geodetic, Lense-Thirring and Thomas precessions (recall that in a
kinematically nonrotating local reference system $R^a_{\
i}=\delta^{aj}$ and $C_a$ represent just the rotation of the local
reference system relative to the global one due to geodetic precession
(see, (\ref{Rij-Ca})).  If, however, a local dynamically nonrotating
reference system is chosen then $C_a=0$ and the term vanishes. Thus, we
prove that in the PPN formalism an extended massive body with mass and
spin only undergoes the same relativistic precession in the background
gravitational field as test massless gyroscope.  This generalizes the
corresponding result for General Relativity given in \cite{dsx:93}.

\section{Outlook and conclusions}
\label{Conclusions-and-outlook}

In this paper we have formulated a new framework of relativistic
celestial mechanics in the first post-Newtonian approximation with PPN
parameters $\gamma$ and $\beta$. Here we list some of the most
interesting results that were derived in the main part of the paper.

\begin{itemize}

\item It is impossible to construct a local reference system satisfying
simultaneously properties {\bf A} and {\bf B} formulated in Section
\ref{Introduction} unless $\eta=0$. Either property {\bf A} or
property {\bf B} separately can be satisfied. For practical
applications the local reference system where the external
gravitational potential is represented by tidal effects only (i.e.,
property {\bf A} is satisfied) seems to be more useful.

\item  As in General Relativity it is still possible to formulate the
theory in terms of two material variables $\sigma$ and $\sigma^i$
related with $T^{\alpha\beta}$ by (\ref{sigma})--(\ref{sigmai}).

\item It is possible to find an empirical definition of mass multipole
moments (\ref{ML-BD-PPN})--(\ref{SL-BD-PPN}) that allows us to keep the
same almost Newtonian multipole expansion (\ref{BD-expansion-PPN}) of
the local gravitational potential $W$ as in General Relativity.  Those
multipole moments generalize the Blanchet-Damour moments introduced in
\cite{blan:damour:89} in the framework of General Relativity.

\item Unlike in General Relativity the BD-like mass and dipole moments
(gravitational mass and mass dipole) do not coincide with the Tolman
(or ADM) mass and dipole moment (inertial mass and mass dipole) even
for one isolated body, where the BD-like mass in general is not
conserved.

\item The tidal expansions of external potentials in the local
reference system involve $\beta$ and $\gamma$. All tidal terms can be
divided into two groups: those which exist in General Relativity
(coefficients now are polynomials of $\beta$ and $\gamma$ rather than
rational numbers) and those vanishing in General Relativity.

\item Equations of motion of a test particle in the local reference
system contain terms proportional to the acceleration of the central
body relative to the global PPN reference system. The effect is
proportional to $\eta$ and is a consequence of a violation of the
Strong Equivalence Principle.

\item For $\eta\neq0$ the equations of motion of a test particle in the
local reference system contain an additional family of multipole
moments ${\cal N}_L$ defined by (\ref{NL}). Those moments also
characterize the gravitational field of the central body, but do not
appear in the multipole expansions
(\ref{BD-expansion-PPN})--(\ref{BD-expansion-PPN-Wi}) of post-Newtonian
gravitational potentials $W$ and $W_a$.

\item For $\beta$ and $\gamma$ unequal to 1 equations of motion of both
test particles (satellites) and massive bodies involve many additional
terms proportional to $\eta$, $\gamma-1$ and $\beta-1$ which vanish in
General Relativity. The terms proportional to $\eta$ come from a
violation of the Strong Equivalence Principle, some other terms result from
a coupling to the external Ricci tensor.

\item It is no longer possible to construct a ``point-mass''
approximation to the equations of motion of $N$ extended bodies by
assuming that all the bodies have constant BD mass and all other BD
mass and spin multipole moments vanish identically as was the case in
General Relativity \cite{dsx:91}. A number of additional assumptions
concerning the moments ${\cal N}_L$ as well as ${\cal P}_a$ and
$\Omega_a$ is indispensable.

\item If $\beta$ and $\gamma$ are unequal to 1 local equations of motions
cannot be represented as a bilinear series of BD-like mass and spin
multipole and tidal moments: one more family of multipole
moments ${\cal N}_L$ and a number of additional terms (containing
${\cal P}_a$, $\Omega$ and $\Omega^a$) appear here.

\item It is possible to derive equations of motion of $N$ massive
extended bodies possessing only mass monopoles ${\cal M}$ (and moments
of inertia ${\cal N}$), and, possibly, spin dipoles ${\cal S}^i$,
without assuming secular stationarity. These equations of motion do not
assume the gravitational masses of the bodies to be constant. The
equations coincide with the known results (e.g., the EIH equations of
motion) in the corresponding limits.

\item Massive bodies possessing only mass monopoles ${\cal M}$ (and
moments of inertia ${\cal N}$) no longer move along a geodesic of the
background metric. This is usually called Nordtvedt effect and presents
a direct consequence of a violation of the Strong Equivalence
Principle. As compared with standard treatment of the Nordtvedt effect
\cite{will:93}, the geodetic deviation of such a body contains not only
terms $\eta\,\Omega$ proportional to internal gravitational energy of
the bodies, but the complete differences between the gravitational and
inertial masses of the bodies (\ref{Delta-M}).

\item Rotational equations of motion contain a term analogous to the
Nordtvedt term in the translational equations of motion (i.e., a term
proportional to $\eta$ and depending on the acceleration of the body
relative to the global PPN metric).  This question is discussed  in
\cite{klioner:soffel:1998} in more detail.

\end{itemize}

The framework presented here extends the Brumberg-Kopeikin and
DSX-formalisms by the introduction of PPN-parameters. It also extends
the classical Nordtvedt-Will PPN-framework by: the formulation of a
theory of astronomical reference frames with new transformation rules
between global and local coordinates comoving with a body of the
gravitational $N$-body system, improved definitions of mass and spin
multipole moments that characterize the gravitational  field of a body
outside of its matter distribution in its own local coordinate system,
and improved translational and rotational equations of motion for a
system of $N$-extended rotating gravitationally interacting bodies of
arbitrary shape and composition. This framework, in our opinion, shifts
the 'pseudo Newtonian view' of the classical PPN-framework towards a
more relativistic one. Note, that this point already was the motivation
for the DSX-formalism.  All main results from the classical
PPN-framework as mentioned in the introduction (Section
\ref{Introduction}) are recovered, but now with the help of an improved
and more consistent framework. Not only the usual PPN
Einstein-Infeld-Hoffmann equations of motion for a system of mass
monopoles, the spin-orbit and spin-spin interaction terms have been
derived in the new $\beta,\gamma$-framework, but also all the various
terms (depending upon $\eta, (\gamma - 1)$ or $(\beta - 1)$) related
with a violation of the Strong Equivalence Principle are recovered.

It is the hope that this version of a PPN-formalism might eventually
form the basis of a new test-theory for experimental gravity. To this
end more work has to be done mainly concerning: a discussion of orders
of magnitude of the various terms that appear in expansions in many
places, the construction of models for the astronomical bodies, e.g.,
for the determination of the time dependence of the various mass and
spin multipole moments of the bodies and, last not least, the relation
of formalism with observations. This last point does not only concern
the theoretical formulation of the various relevant measuring
techniques (VLBI, SLR, LLR, GPS, astrometry, etc.) by employing the new
PPN-framework, but the usefulness of certain theoretical concepts
introduced here, e.g.,  the post-Newtonian spin of a body, has to be
discussed.  It is obvious that one wants to keep concepts that have
been extremely useful in Newtonian classical mechanics for reasons of
simplification also in a relativistic framework. However, the fact that
many of these concepts do not even have a meaning in GRT (without
resorting to some approximations) for real astronomical situations
sheds some light on the difficulties that one faces here.  On the other
hand it is also obvious that the relevant measuring techniques at least
in certain fields of application such as the problem of Earth's
rotation have reached an accuracy level where a wealth of physical
effects enters that simply cannot be modeled.  In that case it has
become a practice to start with a theoretical 'approximate model' and
to measure 'offsets' from the theoretical predictions. What parts of
the new PPN-framework presented here will eventually play a role in
such an 'approximate model' (e.g., for the description of Earth's
rotation) in order to reduce residuals is not so clear at present.

%%% preprint
\addcontentsline{toc}{section}{Acknowledgments}

\acknowledgements

We are grateful to Prof. K. Nordtvedt for fruitful discussions.  S.K.
is also thankful to Dr. S.M. Kopeikin for a number of interesting
discussions raising his interest to the problem of local reference
systems in the PPN formalism.

\end{document}